\documentclass[aps,prd,10pt,showpacs,amsmath,showkeys,twocolumn,floatfix,amssymb, preprintnumbers, nofootinbib, superscriptaddress]{revtex4-1} 
\usepackage{epsfig,dcolumn}
\usepackage{graphicx}
\usepackage{comment} 
\usepackage[usenames]{color}
\usepackage{graphicx}
\usepackage{bm}
 \usepackage{ifpdf}
  \usepackage{floatrow}
\usepackage{makecell}
 \usepackage{caption}

\usepackage[normalem]{ulem}
\usepackage[dvipsnames]{xcolor}
\usepackage[utf8]{inputenc}
\usepackage{hyperref}
\hypersetup{ 
  pdfnewwindow=true,      
  colorlinks=true,        
  linkcolor=PineGreen,    
  citecolor=PineGreen,    
  filecolor=PineGreen,    
  urlcolor=PineGreen      
}

\newcommand{\be}{\begin{eqnarray}}
\newcommand{\ee}{\end{eqnarray}}

 \usepackage{epsfig,dcolumn}
\usepackage{graphicx}
\usepackage{comment} 
\usepackage[usenames]{color}
\usepackage{bm}
\usepackage{ifpdf}
\usepackage{floatrow}
\usepackage{makecell}
\usepackage{caption}
 
\begin{document}

\title{Charged particles interaction in both a finite volume and a uniform magnetic field}

\author{Peng~Guo}
\email{pguo@csub.edu}

\affiliation{Department of Physics and Engineering,  California State University, Bakersfield, CA 93311, USA}
\affiliation{Kavli Institute for Theoretical Physics, University of California, Santa Barbara, CA 93106, USA}

\author{Vladimir~Gasparian}
\affiliation{Department of Physics and Engineering,  California State University, Bakersfield, CA 93311, USA}

\date{\today}

\begin{abstract} 
A formalism for describing charged particles interaction in both a finite volume and  a uniform magnetic field is presented. In the case of    short-range   interaction between charged particles,  we show that  the factorization between short-range physics and finite volume long-range correlation  effect   is possible,  a   L\"uscher formula-like quantization condition is thus obtained.
\end{abstract}

\maketitle

\section{Introduction}\label{intro}

In recent years, a great effort in   nuclear and hadron physics community has been put into constructing the scattering dynamics of few-particle interactions from discrete bound state energy spectrum that is computed in various type of traps, such as commonly used periodic finite box in lattice QCD (LQCD) and harmonic oscillator trap  in nuclear physics computation. The ultimate goal is of course to study and explore the nature of   particle interactions that  plays an  essential role in many fields of physical science, such as   nuclear physics and astrophysics. However, the current state-of-art {\it ab initio} computations in   nuclear   and hadron physics are normally performed in a harmonic oscillator trap and in a  finite volume respectively.   Instead of computing   few-body scattering amplitudes,    the discrete bound state energy levels  are usually  directly measured and extracted from   these  {\it ab initio} computations. Therefore, finding a relation that convert discrete bound state energy spectrum into continuum scattering state is a key step.

In fact, relating the energy shift caused by particle interactions  to the on-shell scattering parameters such as phase shift has the long history across   many fields in physics. In  general cases, the dynamics of particles interaction in a traps is associated to the infinite volume off-shell reaction amplitudes in a highly non-trivial way. Fortunately, when the separation of two physics scales, the size of trap and the range of particles interaction, is clearly established, a simple asymptotic form can be found, which provides a   relation between energy levels in a trap and infinite volume scattering phase shift.
In finite volume in LQCD computation, such a relation in elastic two-body sector is known as L\"uscher formula  \cite{Luscher:1990ux}, which shows a clear factorization of short-range dynamics  and long-range correlation effects because of periodic boundary condition. The short-range dynamics and long-range correlations   are described by  the physical scattering phase shift and L\"uscher's zeta function respectively.    L\"uscher formula has been proving  very successful in LQCD community,  and it has been quickly extended into both coupled-channel and  few-body sectors, see \cite{Rummukainen:1995vs,Christ:2005gi,Bernard:2008ax,He:2005ey,Lage:2009zv,Doring:2011vk,Guo:2012hv,Guo:2013vsa,Kreuzer:2008bi,Polejaeva:2012ut,Hansen:2014eka,Mai:2017bge,Mai:2018djl,Doring:2018xxx,Guo:2016fgl,Guo:2017ism,Guo:2017crd,Guo:2018xbv,Mai:2019fba,Guo:2018ibd,Guo:2019hih,Guo:2019ogp,Guo:2020wbl,Guo:2020kph,Guo:2020iep,Guo:2020ikh,Guo:2020spn,Aoki:2007rd,Feng:2010es,Lang:2011mn,Aoki:2011yj,Dudek:2012gj,Wilson:2014cna,Beane:2007es, Detmold:2008fn, Horz:2019rrn,Brett:2021wyd,Alexandru:2020xqf}. 
In nuclear physics where  a harmonic oscillator trap is commonly used, such a relation is given by BERW formula \cite{Busch98,Stetcu:2007ms, Stetcu:2010xq, Rotureau:2010uz,Rotureau:2011vf,Luu:2010hw,Yang:2016brl,Johnson:2019sps,Zhang:2019cai, Zhang:2020rhz}. In addition to periodic boundary condition and harmonic trap, other type of traps or boundary conditions are also commonly used in different physics fields,   such as hard wall trap \cite{Elhatisari:2016hby,Rokash:2015hra}. 
Regardless difference among various traps, the same strategy is shared: as the two physical scales are clearly separated, a closed asymptotic form can be found, in which short-range dynamics is described by scattering phase shift and long-range   effect is  given by an analytic form that describes how the propagation of particles is affected by the trap, e.g. L\"uscher's zeta function in periodic boundary condition.

In present work,  we aim to establish a similar  relation to  L\"uscher and BERW formula for the charged particles interacting in both a uniform magnetic field and  a periodic box. We remark that   only short-range interaction which represents nuclear force or hadron interactions is considered in this work, the Coulomb interaction has not been incorporated in current framework yet. We also emphasis that   the   Coulomb repulsion   may be important near threshold \cite{Kong:1999sf,Beane:2020ycc,Beane:2014qha,Stellin:2020gst},  especially the long-range nature of Coulomb interaction may complicate the factorization  of physics at different scales and distort the asymptotic wave functions,    the Coulomb interaction  must be included in future work.  We will show that with only a short-range potential, the factorization of short-range physics and long-range correlation effect is possible. Hence a  relation in a compact form that relates discrete energy spectrum to scattering phase shifts can be found.  Such a relation may be useful for the study of charged hadron system such as $\pi^+$ system in LQCD computation.  In finite volume, in order to preserve translation symmetry of system in magnetic field, the magnetic flux though perpendicular  surface of cubic box to a uniform magnetic field must be $2\pi$ multiplied by a rational number $n_p/n_q$, where $n_p$ and $n_q$ are integers and relatively prime  to each other. Therefore, the original energy level without magnetic field is split into $n_q$ sublevels due to the application of magnetic field.  We also remark that the ultimate goal of the current work is to set up a foundation for  exploring the possibility of the topological edge type states \cite{PhysRevLett.49.405,PhysRevLett.71.3697,Hatsugai_1997} in lattice QCD in future work. However, by using  background-field methods in lattice QCD \cite{Detmold:2004qn,Detmold:2004kw,Detmold:2008xk,Detmold:2009fr}, the finite volume energy levels of particle interacting in magnetic field background may also be used to determine the coefficient of the leading local four-nucleon operator contributing to the neutral- and charged-current break-up of the deuteron.

The paper is organized as follows.  The  general  formalism of charged bosons interaction in both a finite volume and  a uniform magnetic field  is   presented  in details  in Section \ref{formalism}. The $S$-wave contribution and regularization of ultraviolet divergence  are discussed    in Section \ref{swave}.   A summary     is given  in Section  \ref{summ}.

\section{Finite volume dynamics of charged bosons in a uniform magnetic field}\label{formalism}

In this section, we briefly summerize the dynamics of charged bosons interacting in both a finite periodic box and   a uniform magnetic field. The uniform magnetic field is chosen   along $z$-axis, $\mathbf{ B}= B \mathbf{ e}_z$, and Landau gauge for vector potential is adopted in this work,
\begin{equation}
\mathbf{ A} (\mathbf{ x} ) = B (0, x, 0).
\end{equation} 
The complete presentation and more rigorous discussion are given in Appendix~\ref{appenddynamics}   and \ref{trapscatt}.

The  dynamics of relative motion    of two charged identical non-relativistic spinless   particles in a uniform magnetic field is described  by    Schr\"odinger equation,
 \begin{equation}
 \left (  \hat{H}_\mathbf{ r}+ V(\mathbf{ r} )\right ) \psi_\varepsilon(\mathbf{ r} ) =\varepsilon \psi_\varepsilon(\mathbf{ r} ), \label{relschrodinger}
 \end{equation}
where $ \psi_\varepsilon(\mathbf{ r} )$ and $\varepsilon$ are wave function and energy for relative motion of two charged boson system. The Hamiltonian operator $ \hat{H}_\mathbf{ r} $ is defined by
 \begin{equation}
  \hat{H}_\mathbf{ r} = -   \frac{\left (\nabla_\mathbf{ r}+ i q \mathbf{ A} (\mathbf{ r}) \right )^2}{2\mu}  ,
 \end{equation}
 where $\mu$ and  $q$  denote reduced mass and charge of two charged particles respectively.

\subsection{Magnetic periodic boundary condition} 
 In a periodic finite box, though the short-range potential $V$ is periodic 
 \begin{equation}
 V(\mathbf{ r} + \mathbf{ n} L) =  V(\mathbf{ r}  ), \ \ \ \ \mathbf{ n} \in \mathbb{Z}^3,
 \end{equation}
 where $L$ denotes the size of box, the    $\hat{H}_\mathbf{ r} $ is not discrete translation invariant  
 $$\hat{H}_{\mathbf{ r}  + \mathbf{ n} L}\neq \hat{H}_\mathbf{ r}$$
  due to the fact that    vector potential is  coordinate dependent, and breaks discrete translation symmetry  
  $$A_y (\mathbf{ r} + \mathbf{ n} L) = A_y (\mathbf{ r}) + B\mathbf{ n} L \cdot \mathbf{ e}_x .$$ The momentum operator, $\mathbf{ \hat{p}} = - i \nabla_\mathbf{ r}$, doesn't commute with $\hat{H}_\mathbf{ r} $: $$[\mathbf{ \hat{p}}, \hat{H}_\mathbf{ r} ] \neq 0. $$ Hence canonical momentum is no longer a conserved quantity as the consequence of breaking down of discrete translation symmetry in a uniform magnetic field. It has been shown in Refs.~\cite{PhysRev.133.A1038,PhysRev.134.A1602,yoshioka2002quantum} that a pseudo-momentum operator
  \begin{equation}
  \mathbf{ \hat{K}}_\mathbf{ r} = - i \nabla_\mathbf{ r} + q \mathbf{ A} (\mathbf{ r} )  - q \mathbf{ B}  \times \mathbf{ r}= - i \nabla_\mathbf{ r} + qB (r_y, 0,0)
  \end{equation}
in fact commute with $\hat{H}_\mathbf{ r} $: $$[\mathbf{ \hat{K}}_\mathbf{ r}, \hat{H}_\mathbf{ r} ] = 0.$$ Therefore,    $\mathbf{ \hat{K}}_\mathbf{ r}$ can be used as generator of a magnetic translation operator,
\begin{equation}
\hat{T}_\mathbf{ r} (\mathbf{ n} L)  = e^{ i \mathbf{ \hat{K}}_\mathbf{ r} \cdot \mathbf{ n} L },  
\end{equation}
and
\begin{equation}
  [ \hat{T}_\mathbf{ r} (\mathbf{ n} L)  , \hat{H}_\mathbf{ r} ] = 0, \ \ \ \  \mathbf{ n} \in \mathbb{Z}^3.
\end{equation}

However, the magnetic translation operator in general doesn't commute with each other, for instance, for a closed   path in a single box,
\begin{equation}
\hat{T}_\mathbf{ r} ( - L \mathbf{ e}_x)  \hat{T}_\mathbf{ r} ( - L \mathbf{ e}_y) \hat{T}_\mathbf{ r} (  L \mathbf{ e}_x)  \hat{T}_\mathbf{ r} (  L \mathbf{ e}_y)  =e^{- i q B L^2}  \neq 1.
\end{equation}
 To warrant a state that is translated through a closed path remain same, the magnetic flux $qB L^2$ through the surface of path must be quantized,
 $$q B L^2 = 2\pi n, \ \  n \in \mathbb{Z}.$$ In fact, this conclusion can be made in a more general way by considering a enlarged closed path in $x-y$ plane with the size of $n_q L\mathbf{  e}_x \times L \mathbf{ e}_y$ where $n_q \in \mathbb{Z}$, see  Refs.~\cite{PhysRev.133.A1038,PhysRev.134.A1602}.  Hence, the generalized magnetic quantization condition is given by
 \begin{equation}
 q B n_q L^2 = 2\pi  n_p , 
 \end{equation}
 where $n_p$ and $ n_q$ are two relatively prime integers. In a enlarged magnetic unit box defined by magnetic unit vectors: $$n_q L\mathbf{  e}_x \times L \mathbf{ e}_y\times L \mathbf{ e}_z,$$  the magnetic translation operators now commute with each other 
  \begin{equation}
[  \hat{T}_\mathbf{ r} (  n_q L \mathbf{ e}_x) ,   \hat{T}_\mathbf{ r} (  L \mathbf{ e}_y) ] = 0.
 \end{equation}
 Therefore,  the discrete translation in a enlarged magnetic unit box leaves Hamiltonian     invariant, and $\hat{T}_\mathbf{ r} (\mathbf{ n}_B L) $   form a magnetic translation group, where
 \begin{equation}
 \mathbf{ n}_B = n_x n_q  \mathbf{e }_x + n_y \mathbf{ e}_y + n_z \mathbf{ e}_z, \ \ \ \  n_{x,y,z} \in \mathbb{Z}.
 \end{equation}
  The application of the magnetic field also results in the splitting of each energy level    into $n_q$ sub-energy levels.

 Under magnetic translation operation, the wave function  behaves as
 \begin{equation}
 \hat{T}_\mathbf{ r} (\mathbf{ n}_B L)   \psi_\varepsilon(\mathbf{ r} ) =e^{i q B n_x n_q L r_y}  \psi_\varepsilon(\mathbf{ r} + \mathbf{ n}_B L ).
 \end{equation}
 According to Bloch theorem, in a periodic box, periodicity of system requires that $ \hat{T}_\mathbf{ r} (\mathbf{ n}_B L)   \psi_\varepsilon(\mathbf{ r} )$ can  only differ from $   \psi_\varepsilon(\mathbf{ r} )$ by a phase factor, which can be chosen as $$ e^{ i \frac{\mathbf{ P}_B}{2} \cdot \mathbf{ n}_B L} ,$$ where 
  \begin{equation}
 \mathbf{ P}_B =  \frac{2\pi}{L} \left ( \frac{n_x}{ n_q}  \mathbf{e }_x + n_y \mathbf{ e}_y + n_z \mathbf{ e}_z \right ), \ \ \ \  n_{x,y,z} \in \mathbb{Z}.
 \end{equation}
 Hence the magnetic periodic boundary condition is given by
 \begin{equation}
 \psi_\varepsilon(\mathbf{ r} + \mathbf{ n}_B L ) = e^{ i \frac{\mathbf{ P}_B}{2} \cdot \mathbf{ n}_B L}  e^{-i q B n_x n_q L r_y}   \psi_\varepsilon(\mathbf{ r} ). \label{magneticperiodic}
 \end{equation}
 The  magnetic periodic boundary condition can also be obtained by considering separable form of total wave function, see  Appendix~\ref{MTranslation}.

\subsection{Finite volume Lippmann-Schwinger equation and quantization condition} 

\subsubsection{Finite volume Lippmann-Schwinger equation}
The   Schr\"odinger equation  and magnetic periodic  boundary condition in Eq.(\ref{relschrodinger}) and Eq.(\ref{magneticperiodic})   together  can be replaced   by finite volume  homogeneous Lippmann-Schwinger (LS) equation,
\begin{equation}
\psi_{  \varepsilon}(\mathbf{ r})  = \int_{L_B^3} d \mathbf{ r}' G^{(L)}_B(\mathbf{ r},\mathbf{ r}'; \varepsilon) V(\mathbf{ r}')\psi_{  \varepsilon}(\mathbf{ r}')  , \label{homogenousLSB}
\end{equation}
 where  the volume integration over the magnetic unit cell is defined by
 \begin{equation}
  \int_{L_B^3} d \mathbf{ r}'  = \int_{- \frac{n_q L}{2}}^{\frac{n_q L}{2}}  d r_x'   \int_{- \frac{  L}{2}}^{\frac{  L}{2}}  d r'_y  \int_{- \frac{  L}{2}}^{\frac{  L}{2}} d r'_z.
  \end{equation}
 The finite volume magnetic Green's function $G^{(L)}_B$ also must     satisfies the   magnetic periodic boundary condition,
 \begin{align}
& G^{(L)}_B(\mathbf{ r},\mathbf{ r}'; \varepsilon) \nonumber \\
& = e^{ - i    \frac{ \mathbf{ P}_B  }{2}    \cdot  \mathbf{ n}_B L }   e^{  i   q B n_x n_q L r_y   } G^{(L)}_B(\mathbf{ r}+  \mathbf{ n}_B  L,\mathbf{ r}'; \varepsilon)   \nonumber \\
 &= e^{  i    \frac{ \mathbf{ P}_B  }{2}    \cdot  \mathbf{ n}_B L }   e^{ - i   q B n_x n_q L r'_y   }  G^{(L)}_B(\mathbf{ r},\mathbf{ r}'+  \mathbf{ n}_B  L; \varepsilon) , 
 \end{align}
hence dynamical equation for  $G^{(L)}_B$  is given by
 \begin{align}
& \left ( \varepsilon -  \hat{H}_\mathbf{ r}   \right ) G^{(L)}_B(\mathbf{ r},\mathbf{ r}'; \varepsilon)  \nonumber \\
&= \sum_{\mathbf{ n}_B}  e^{  -i    \frac{ \mathbf{ P}_B  }{2}    \cdot  \mathbf{ n}_B L }  e^{  i   q B r_y \mathbf{ e}_x   \cdot  \mathbf{ n}_B L }  \delta(\mathbf{ r}-\mathbf{ r}' + \mathbf{ n}_B L). \label{magneticGBeq}
 \end{align}

The   solution of finite vollume magnetic Green's function $G^{(L)}_B$   can be constructed from its infinite volume counterpart $G^{(\infty)}_B$ by, 
\begin{align}
& G^{(L)}_B(\mathbf{ r},\mathbf{ r}'; \varepsilon) \nonumber \\
& = \sum_{\mathbf{ n}_B}    G^{(\infty)}_B(\mathbf{ r},\mathbf{ r}' + \mathbf{ n}_B L; \varepsilon)   e^{  i    \frac{ \mathbf{ P}_B  }{2}    \cdot  \mathbf{ n}_B L }   e^{  -i   q B r'_y \mathbf{ e}_x   \cdot  \mathbf{ n}_B L }  \nonumber \\
& = \sum_{\mathbf{ n}_B}     e^{ - i    \frac{ \mathbf{ P}_B  }{2}    \cdot  \mathbf{ n}_B L }   e^{  i   q B r_y \mathbf{ e}_x   \cdot  \mathbf{ n}_B L } G^{(\infty)}_B(\mathbf{ r} + \mathbf{ n}_B L,\mathbf{ r}' ; \varepsilon)  , 
\end{align}
  details of construction can be found   in Appendix~\ref{relativeLSeq}.
The infinite volume magnetic Green's function $G^{(\infty)}_B$ satisfies equation,
  \begin{equation}
 \left ( \varepsilon -  \hat{H}_\mathbf{ r}   \right ) G^{(\infty)}_B(\mathbf{ r},\mathbf{ r}'; \varepsilon)  =  \delta(\mathbf{ r}-\mathbf{ r}'  ),
 \end{equation}
 and the analytic expression of $G^{(\infty)}_B$ is given by
 \begin{align}
 & G^{(\infty)}_B(\mathbf{ r},\mathbf{ r}'; \varepsilon)    = -  \frac{2\mu qB}{2\pi}  e^{ - \frac{i q B}{2} (r_x + r'_x) (r_y - r'_y)} e^{- \frac{qB}{4}  | \bm{\rho} - \bm{\rho}' |^2  }   \nonumber \\
 & \times  \sum_{n=0}^\infty \frac{i L_n ( \frac{qB}{2}  | \bm{\rho} - \bm{\rho}' |^2 ) e^{i \sqrt{2\mu \varepsilon - 2qB (n+ \frac{1}{2})  } |r_z - r'_z|}}{2 \sqrt{2\mu \varepsilon - 2qB (n+ \frac{1}{2})  }}  ,
 \end{align}
where $L_n (x)$ is Laguerre polynomial,  and $$\bm{\rho} = r_x \mathbf{ e}_x + r_y \mathbf{ e}_y, \ \ \ \ \bm{\rho}' = r'_x \mathbf{ e}_x + r'_y \mathbf{ e}_y$$ are relative coordinates defined in $x-y$ plane.

\subsubsection{Quantization condition with short-range interaction}
The discrete bound state energy spectrum can be found as the eigen-energy solutions of homogeneous LS equation in Eq.(\ref{homogenousLSB}). The partial wave expansion in angular momentum basis is commonly used in describing   infinite volume scattering state. However in magnetic field, due to asymmetry of magnetic Hamiltonian in $x-y$ plane and along $z$-axis,    angular momentum basis in spherical coordinates  is in fact not most convenient basis in describing   dynamics of charged particles in uniform magnetic field. Nevertheless, it can be done in principle. For the sake of   the consistency of presentation in both finite volume and infinite volume dynamics. Let's consider the partial wave expansion of Eq.(\ref{homogenousLSB}). Using
\begin{equation}
\psi_{  \varepsilon}(\mathbf{ r}) = \sum_{lm} \psi_{  l m}^{(L)}(r) Y_{lm } (\mathbf{ \hat{r}}),
\end{equation}
and
\begin{equation}
G^{(L)}_B(\mathbf{ r},\mathbf{ r}'; \varepsilon)   = \sum_{lm, l'm'}  Y_{lm } (\mathbf{ \hat{r}}) G^{(B, L)}_{lm, l'm'}(r,r'; \varepsilon)  Y^*_{l'm' } (\mathbf{ \hat{r}}') ,
\end{equation}
 we   find
\begin{equation}
 \psi^{(L)}_{lm }(r)  = \sum_{l' m'} \int_{L_B^3} {r'}^2 d r' G_{ lm, l'm'}^{(B,L)} (r, r' ; \varepsilon)  V_{l'} ( r')   \psi^{(L)}_{l'm'}(r')   . \label{homogeneousLSPWA}
\end{equation}

Since the purpose of this work is to find a  L\"uscher formula-like simple relation that connects short-range physics associated to particles interaction  $V(\mathbf{ r})$ and the long-range effect generated by the finite volume and magnetic field. Also considering the fact that such a relation is result of clear separation of two    physical scales: (1) the range of potential $V(\mathbf{ r})$ and (2) the size of a trap or finite volume. When the two scales are clearly separated, the short- and long-range physics can be factorized, and a compact relation as the leading order contribution can be found  by studying the asymptotic behavior of wave function \cite{Guo:2012hv,Guo:2019hih}. Therefore, for our purpose, it is sufficient to consider zero-range potential,   
 \begin{equation}
 V_{l}(r) \rightarrow V_l \frac{ \delta(r)  }{ r^2}     \frac{2^{2l+1} \Gamma^2 (l+ \frac{3}{2})}{ (2\pi)^3  r^{2l} }  , \label{zeroVpot}
 \end{equation}
see Appendix~\ref{trapscatt} for the more rigorous discussion.
 The Eq.(\ref{homogeneousLSPWA}) is thus turned into an algebra equation,
 \begin{align}
 \frac{\psi^{(L)}_{lm }(r)}{r^l}  &= \sum_{l' m'}  V_{l'}   \frac{2^{2l'+1} \Gamma^2 (l'+ \frac{3}{2})}{(2\pi)^3}     \nonumber \\
 & \times     \frac{G_{ lm, l'm'}^{(B,L)} (r, r' ; \varepsilon) }{   r^l  {r' }^{l'} }  \frac{   \psi^{(L)}_{l'm'}(r') }{{r'}^{l'}}|_{r' \rightarrow 0}   .
\end{align}  
Hence the quantization condition of discrete energy spectrum   is given  by
 \begin{equation}
  \det   \left [  \frac{\delta_{lm, l'm'}}{2^{2l+1} V_l} -   \frac{ \Gamma^2 (l'+ \frac{3}{2})}{(2\pi)^3}   \frac{G_{ lm, l'm'}^{(B,L)} (r, r' ; \varepsilon) }{   r^l  {r' }^{l'} } |_{r,r'\rightarrow 0}  \right ]   =0 . \label{magneticQCV}
\end{equation}

Under the same assumption of zero-range approximation given in Eq.(\ref{zeroVpot}), the potential strength $V_l$ is related to the infinite volume two-body  scattering phase shift $\delta_l (k_\varepsilon)$ by
\begin{equation}
  \frac{(4\pi)^2}{ 2\mu V_l }+ \frac{2^{2l+1}\Gamma (l+ \frac{1}{2}) \Gamma (l+ \frac{3}{2}) }{ \pi   r^{2l+1}}   |_{r \rightarrow 0}  =  -     k_\varepsilon^{2 l+1}     \cot  \delta_l(k_\varepsilon)      ,   \label{phaseshiftinf}
\end{equation} 
  see detailed discussions in Appendix~\ref{trapscatt}. The relative momentum $k_\varepsilon$ in infinite volume is related to the relative finite volume energy $\varepsilon$ by
 \begin{equation}
 \frac{k_\varepsilon^2}{2\mu} = \varepsilon + \triangle E_{R}, \ \ \ \  \triangle E_{R}= \frac{QB}{M} (n+ \frac{1}{2} ) - \frac{ P_x^2 + P_y^2}{2M} ,
 \end{equation} 
 where $\triangle E_{R}$ is the result of quantization of CM motion in uniform magnetic field.

Eliminating $V_l$, the   Eq.(\ref{magneticQCV}) and Eq.(\ref{phaseshiftinf})  together   yield a L\"uscher formula-like simple relation,
 \begin{equation}
  \det   \left  [  \delta_{lm, l'm'}        \cot \delta_l (k_\varepsilon)  - \mathcal{M}^{(B,L)}_{lm, l'm'}  (\varepsilon)    \right  ]   =0 ,
\end{equation}  
where
 \begin{align}
  &  \mathcal{M}^{(B,L)}_{lm, l'm'}  (\varepsilon)   =- \frac{2^{2l'+3} \Gamma^2 (l'+ \frac{3}{2})}{ 2\mu k_\varepsilon^{2l+1}  (2\pi)}   \frac{G_{ lm, l'm'}^{(B,L)} (r, r' ; \varepsilon) }{   r^l  {r' }^{l'} } |_{r,r'\rightarrow 0}  \nonumber \\
  &  - \delta_{lm, l'm'}   \frac{2^{2l+1} \Gamma(l+\frac{1}{2})\Gamma(l+\frac{3}{2})}{\pi} \frac{1}{(k_\varepsilon r)^{2l+1}} |_{r\rightarrow 0}   .  
\end{align}  
The second term in $ \mathcal{M}^{(B,L)}_{lm, l'm'} $ plays the role of the regulator of ultraviolet (UV) divergence  and will cancel out the UV divergence in finite volume magnetic Green's function, so that $ \mathcal{M}^{(B,L)}_{lm, l'm'} $ is ultimately free of UV divergence. In general, the regularization and isolation of  UV divergence in higher partial waves of finite volume magnetic Green's function is a highly non-trivial task. Fortunately, it can be accomplished rather neatly for $S$-wave, hence, only $S$-wave contribution will be considered in   Section \ref{swave}. The regularization of UV divergence will be worked out explicitly.

\section{$S$-wave contribution and contact interaction}\label{swave}
As already mentioned in previous section, the angular momentum basis in general is not convenient basis for the dynamics of charged particles in uniform magnetic field. The partial wave expansion of finite volume magnetic Green's function and ultraviolet regularization can be tedious in general. Fortunately, if only $S$-wave contribution is dominant, the formalism can be worked out nicely. In this section, only a contact interaction potential
\begin{equation}
V(\mathbf{ r}) = \frac{V_0}{4\pi} \delta(\mathbf{ r})
\end{equation}
is used, which may be considered as the leading order contribution of chiral effective field theory and may be suitable for the few-body system, such as $\pi^+$ interactions in finite volume.

With a contact interaction, the finite volume quantization condition is simply given by
\begin{equation}
\frac{4\pi}{V_0} =  G^{(L)}_B(\mathbf{ 0},\mathbf{ 0}; \varepsilon)  .
\end{equation}
In infinite volume, $V_0$ is related to $S$-wave scattering amplitude by
\begin{equation}
\frac{4\pi}{V_0} - G^{(\infty)} (\mathbf{ 0} ; k_\varepsilon)  =- \frac{2\mu k_\varepsilon}{4\pi} \frac{1}{t_0 (k_\varepsilon)},
\end{equation}
where $$t_0(k_\varepsilon) = \frac{1}{\cot \delta_0(k_\varepsilon) - i},$$  and infinite volume Green's function $G^{(\infty)} (\mathbf{ 0} ; k_\varepsilon) $ is given by
\begin{equation}
G^{(\infty)} (\mathbf{ 0} ; k_\varepsilon)= \int \frac{d \mathbf{ p}}{(2\pi)^3} \frac{ 1 }{\frac{ k_\varepsilon^2}{2\mu} - \frac{\mathbf{ p}^2}{2\mu}}  = -  i  \frac{2 \mu k_\varepsilon}{4\pi}  - \frac{2 \mu}{4\pi r}  |_{r\rightarrow 0} .
\end{equation}
Thus, the quantization condition is simply given by
\begin{equation}
  \cot \delta_0 (k_\varepsilon)= - \frac{4\pi}{2\mu k_\varepsilon}  G^{(L)}_B(\mathbf{ 0},\mathbf{ 0}; \varepsilon)  - \frac{1}{ k_\varepsilon r}  |_{r\rightarrow 0}  . \label{unregQC}
\end{equation}
 The magnetic Green's function $G^{(L)}_B(\mathbf{ 0},\mathbf{ 0}; \varepsilon) $ is a real function of $\varepsilon$. The  UV divergent term $$-\frac{1}{ k_\varepsilon r}  |_{r\rightarrow 0}$$ play the role of UV counter term that cancel out the UV divergent term in $G^{(L)}_B(\mathbf{ 0},\mathbf{ 0}; \varepsilon) $, so ultimate result is finite and real as a  function of $\varepsilon$.

\subsection{Regularization of UV divergence} 
 In this section, we show explicitly how the UV divergence in  $G^{(L)}_B(\mathbf{ 0},\mathbf{ 0}; \varepsilon) $ is regularized and isolated out explicitly. The UV divergence only appear when $$\mathbf{ r}=( \bm{\rho}  , 0) \rightarrow \mathbf{ 0}, $$ hence, a small $  r=\rho$  is used as UV regulator, in the end, final expression is obtained by taking the limit of $r= \rho \rightarrow 0$. Starting with explicit expression of magnetic Green's function in CM frame ($\mathbf{ P}_B =\mathbf{ 0}$), 
  \begin{align}
 & G^{(L)}_B( \mathbf{ 0},\mathbf{ 0}; \varepsilon)  =  \sum_{n_x, n_y \in \mathbb{Z}}        e^{ - \frac{i q B}{2}   n_x n_q L  n_y L} \nonumber \\
 & \times  e^{- \frac{qB}{4}  | \bm{\rho}  + n_x n_q L \mathbf{ e}_x + n_y L \mathbf{ e}_y|^2  }  \frac{qB}{2\pi}   \frac{1}{L} \sum_{  k_{z} = \frac{2\pi n_{z}}{L}  }^{ n_{z} \in \mathbb{Z}}    \nonumber \\
 & \times    \sum_{n=0}^\infty \frac{ L_n ( \frac{qB}{2}  | \bm{\rho}   + n_x n_q L \mathbf{ e}_x + n_y L \mathbf{ e}_y|^2 ) }{\varepsilon - \frac{qB}{\mu}(n+ \frac{1}{2}) - \frac{k_z^2}{2\mu} } |_{\bm{\rho} \rightarrow 0}. \label{GBL00}
 \end{align}
 The UV divergence is associated to the term
   \begin{align}
     & \frac{1}{L} \sum_{  k_{z} = \frac{2\pi n_{z}}{L}  }^{ n_{z} \in \mathbb{Z}}      \sum_{n=0}^\infty \frac{1}{\varepsilon - \frac{qB}{\mu}(n+ \frac{1}{2}) - \frac{k_z^2}{2\mu} }  \nonumber \\
      &\propto  \int^{\Lambda = \frac{1}{r}} \frac{d k^3}{k^2} \propto \Lambda = \frac{1}{r}|_{r\rightarrow 0},
 \end{align}
  hence $G^{(L)}_B$ is linearly divergent.

  The linear divergence can be regularized simply by subtraction. Therefore, we  firstly split finite volume magnetic Green's function   into regularized term by subtraction and a term that is UV divergent,
  \begin{equation}
  G^{(L)}_B( \mathbf{ 0},\mathbf{ 0}; \varepsilon)   = \triangle G^{(L)}_B (\varepsilon) +  G^{(L)}_B(\mathbf{ 0},\mathbf{ 0};  0),
  \end{equation}
 where 
 \begin{equation}
 \triangle G^{(L)}_B(  \varepsilon)  =   G^{(L)}_B(\mathbf{ 0},\mathbf{ 0}; \varepsilon)   - G^{(L)}_B(\mathbf{ 0},\mathbf{ 0}; 0)    .
 \end{equation}
 The subtracted term   $\triangle G^{(L)}_B(  \varepsilon)  $ is free of UV divergence. Using identity
 \begin{equation}
 \frac{1}{L} \sum_{  k_{z} = \frac{2\pi n_{z}}{L}   }^{ n_{z} \in \mathbb{Z}}  \frac{1}{E - \frac{k_z^2}{2\mu}} = \frac{2\mu}{2 \sqrt{2\mu E}} \cot \frac{\sqrt{2\mu E} L}{2}, \label{kzlatticesum}
 \end{equation}
 $\triangle G^{(L)}_B(  \varepsilon)  $ is thus given explicitly by
  \begin{align}
 & \triangle G^{(L)}_B( \varepsilon) =  \frac{2\mu qB}{4\pi}    \sum_{n=0}^\infty   \sum_{n_x, n_y \in \mathbb{Z}}       e^{ - \frac{i q B n_x    n_y  n_q L^2 }{2}   }  \nonumber \\
 & \times    e^{- \frac{qB}{4}  |  n_x n_q L \mathbf{ e}_x + n_y L \mathbf{ e}_y|^2  }    L_n ( \frac{qB}{2}  |   n_x n_q L \mathbf{ e}_x + n_y L \mathbf{ e}_y|^2 )   \nonumber \\
 & \times \left [ \frac{\cot \frac{\sqrt{2\mu \varepsilon - 2qB (n+ \frac{1}{2})  } L}{2}}{\sqrt{2\mu \varepsilon - 2qB (n+ \frac{1}{2})  }} + \frac{\coth \frac{\sqrt{   2qB (n+ \frac{1}{2})  } L}{2}}{  \sqrt{  2qB (n+ \frac{1}{2})  }} \right ]. \label{GBLsubtraction}
 \end{align}

Next, the UV divergence  in  $G^{(L)}_B(\mathbf{ 0},\mathbf{ 0};  0)$ can be isolated out by further split $G^{(L)}_B(\mathbf{ 0},\mathbf{ 0};  0)$    into
 \begin{equation}
  G^{(L)}_B(\mathbf{ 0},\mathbf{ 0};  0) =  G^{(B,L)}_{UV}    + G^{(B,L)}_{RC}  , 
 \end{equation}
 where $ G^{(B,L)}_{UV}$ is  UV divergent  and is given by,
  \begin{equation}
 G^{(B,L)}_{UV}  =   \frac{qB}{2\pi}   \frac{1}{L} \sum_{  k_{z} = \frac{2\pi n_{z}}{L}   }^{ n_{z} \in \mathbb{Z}}     \sum_{n=0}^\infty \frac{ L_n ( \frac{qB}{2}  | \bm{\rho}   |^2 ) }{  - \frac{qB}{\mu}(n+ \frac{1}{2}) - \frac{k_z^2}{2\mu} } |_{\bm{\rho} \rightarrow 0}  . \label{GUV}
 \end{equation}
 The $ G^{(B,L)}_{RC}$ is a regulated constant term, and is defined by
 \begin{align}
 & G^{(B,L)}_{RC} =   \sum_{n_x, n_y \neq 0}       e^{ - \frac{i q B n_x    n_y  n_q L^2 }{2}   }  \frac{qB}{2\pi}   \frac{1}{L} \sum_{  k_{z} = \frac{2\pi n_{z}}{L}   }^{ n_{z} \in \mathbb{Z}}    \nonumber \\
 & \times    \sum_{n=0}^\infty \frac{ e^{- \frac{qB}{4}  |  n_x n_q L \mathbf{ e}_x + n_y L \mathbf{ e}_y|^2  }      L_n ( \frac{qB}{2}  |   n_x n_q L \mathbf{ e}_x + n_y L \mathbf{ e}_y|^2 ) }{  - \frac{qB}{\mu}(n+ \frac{1}{2}) - \frac{k_z^2}{2\mu} }.
 \end{align}

(1) The regulated constant term $ G^{(B,L)}_{RC}$ can be further simplified by using    identity
 \begin{align}
  & \sum_{n=0}^\infty \frac{ L_n ( \frac{qB}{2} x^2 ) }{  - \frac{qB}{\mu}(n+ \frac{1}{2}) - \frac{k_z^2}{2\mu} }  \nonumber \\
   & =  - \frac{\mu }{q B} \Gamma(\frac{1}{2}+ \frac{k_z^2}{2q B})  U ( \frac{1}{2} + \frac{k_z^2}{2q B}, 1, \frac{q B}{2} x^2 ), \label{gammakummer}
 \end{align}
where $U(a,b,z)$ stands for Kummer function, hence, we find
  \begin{align}
 & G^{(B,L)}_{RC}  =  -  \frac{2\mu}{4\pi}  \frac{1}{L} \sum_{  k_{z} = \frac{2\pi n_{z}}{L}   }^{ n_{z} \in \mathbb{Z}}    \Gamma(\frac{1}{2}+ \frac{k_z^2}{2q B})   \nonumber \\
 & \times  \sum_{n_x, n_y \neq 0}       e^{ - \frac{i q B n_x    n_y  n_q L^2 }{2}   }    e^{- \frac{qB}{4}  |  n_x n_q L \mathbf{ e}_x + n_y L \mathbf{ e}_y|^2  }      \nonumber \\
 & \times    U ( \frac{1}{2} + \frac{k_z^2}{2q B}, 1, \frac{q B}{2}  |   n_x n_q L \mathbf{ e}_x + n_y L \mathbf{ e}_y|^2 ). \label{GBLRC}
 \end{align}
 Asymptotically,  Kummer function decay exponentially,
 \begin{align}
 &e^{- \frac{qB}{4} x^2}   \Gamma(\frac{1}{2}+ \frac{k_z^2}{2q B}) U( \frac{1}{2} + \frac{k_z^2}{2q B}, 1, \frac{q B}{2} x^2)   \nonumber \\
 &  \stackrel{ k_z \rightarrow \infty}{\rightarrow } 2 K_0 ( \sqrt{  k_z^2  x^2}),
 \end{align}
 hence   $G^{(B,L)}_{RC}  $ is indeed a well-defined regulated constant.
 
 (2) The explicit expression of UV divergence in $ G^{(B,L)}_{UV}$ can be worked out. First of all, the infinite sum of integer $n \in [0, \infty]$  in Eq.(\ref{GUV}) is split into $n \in [0, n_\Lambda]$ and $n \in [n_\Lambda, \infty]$, where $n_\Lambda$ serves as a cutoff integer and $n_\Lambda \gg 1$. For large $n_\Lambda$, the summation is replaced by integration, hence one can rewrite Eq.(\ref{GUV})  to
   \begin{align}
 G^{(B,L)}_{UV} & =   \frac{qB}{2\pi}   \frac{1}{L} \sum_{  k_{z} = \frac{2\pi n_{z}}{L}   }^{ n_{z} \in \mathbb{Z}}     \sum_{n=0}^{n_\Lambda} \frac{ L_n ( \frac{qB}{2}  | \bm{\rho}   |^2 ) }{  - \frac{qB}{\mu}(n+ \frac{1}{2}) - \frac{k_z^2}{2\mu} } |_{\bm{\rho} \rightarrow 0}  \nonumber \\
 &+  \frac{qB}{2\pi}   \frac{1}{L} \sum_{  k_{z} = \frac{2\pi n_{z}}{L}   }^{ n_{z} \in \mathbb{Z}}     \int_{n_\Lambda}^\infty d n \frac{ L_n ( \frac{qB}{2}  | \bm{\rho}   |^2 ) }{  - \frac{qB}{\mu}(n+ \frac{1}{2}) - \frac{k_z^2}{2\mu} } |_{\bm{\rho} \rightarrow 0} .  
 \end{align}
Let's rewrite it further to
    \begin{align}
& G^{(B,L)}_{UV}  \nonumber \\
& =   \frac{qB}{2\pi}   \frac{1}{L} \sum_{  k_{z} = \frac{2\pi n_{z}}{L}   }^{ n_{z} \in \mathbb{Z}} \left [    \sum_{n=0}^{n_\Lambda }- \int_0^{n_\Lambda} d n \right ] \frac{ L_n ( \frac{qB}{2}  | \bm{\rho}   |^2 ) }{  - \frac{qB}{\mu}(n+ \frac{1}{2}) - \frac{k_z^2}{2\mu} } |_{\bm{\rho} \rightarrow 0}  \nonumber \\
 &+  \frac{qB}{2\pi}   \frac{1}{L} \sum_{  k_{z} = \frac{2\pi n_{z}}{L}   }^{ n_{z} \in \mathbb{Z}}     \int_0^\infty d n \frac{ L_n ( \frac{qB}{2}  | \bm{\rho}   |^2 ) }{  - \frac{qB}{\mu}(n+ \frac{1}{2}) - \frac{k_z^2}{2\mu} } |_{\bm{\rho} \rightarrow 0} .   \label{GUVsub}
 \end{align}
 The first term in Eq.(\ref{GUVsub}) is finite, so   $\bm{\rho} $ can be set to zero safely. In the second term in Eq.(\ref{GUVsub}),   let's rescale integral  dummy variable $d n$ to 
 \begin{equation}
q B \rho^2 d n \rightarrow   \xi d \xi, \ \  qB \rho^2 n \rightarrow \frac{\xi^2}{2} ,  
 \end{equation}
 thus, 
   the   Eq.(\ref{GUV}) is then turned into 
       \begin{align}
 G^{(B,L)}_{UV}  
& =   \frac{qB}{2\pi}   \frac{1}{L} \sum_{  k_{z} = \frac{2\pi n_{z}}{L}   }^{ n_{z} \in \mathbb{Z}} \left [    \sum_{n=0}^{n_\Lambda} - \int_0^{n_\Lambda} d n \right ] \frac{ 1}{  - \frac{qB}{\mu}(n+ \frac{1}{2}) - \frac{k_z^2}{2\mu} }    \nonumber \\
 &+   \frac{1}{2\pi}   \frac{1}{L} \sum_{  k_{z} = \frac{2\pi n_{z}}{L}   }^{ n_{z} \in \mathbb{Z}}     \int_0^\infty \xi d \xi  \frac{ L_{  \frac{\xi^2}{2 q B r^2}} ( \frac{qB r^2}{2}   ) }{  - \frac{\xi^2 }{2 \mu} - \frac{qB r^2 }{2 \mu} - \frac{k_z^2 r^2}{2\mu} } |_{ r \rightarrow 0}  .   
 \end{align}
 Using asymptotic form of Laguerre polynomial,
 \begin{equation}
  L_{  \frac{\xi^2}{2 q B r^2}} ( \frac{qB r^2}{2}   ) \stackrel{ r  \rightarrow 0}{\rightarrow } J_0 (\sqrt{\xi^2}),
 \end{equation}
and identity in Eq.(\ref{kzlatticesum}) again,  
 we find
        \begin{align}
& G^{(B,L)}_{UV} = \triangle G^{(B,L)}_{UV}    \nonumber \\
&    -   \frac{2\mu}{2\pi r}      \int_0^\infty \xi d \xi  J_0 (\sqrt{\xi^2}) \frac{\coth \frac{ \sqrt{ \frac{ \xi^2}{r^2}   +  qB   } L}{2} }{ 2  \sqrt{  \xi^2   +  qB r^2  } } |_{ r \rightarrow 0} ,   
 \end{align}
 where
         \begin{equation}
 \triangle G^{(B,L)}_{UV}  
 = - \frac{2\mu qB}{2\pi}     \left [    \sum_{n=0}^{n_\Lambda} - \int_0^{n_\Lambda} d n \right ]  \frac{\coth \frac{\sqrt{2 q B(n+\frac{1}{2})} L}{2}}{2 \sqrt{ 2 q B (n+\frac{1}{2}) }}     .   \label{deltaGUV}
 \end{equation}
 As $r\rightarrow 0$ and $\frac{1}{r} \rightarrow \infty$, 
 \begin{equation} 
 \coth  z \stackrel{z \rightarrow\infty }{ \sim }1 +2 e^{-2 z} + 2 e^{-4 z} \cdots, \ \  z=\frac{ \sqrt{ \frac{ \xi^2}{r^2}   +  qB   } L}{2} ,
 \end{equation}
and also using identity
 \begin{equation}
  \int_0^\infty \xi d \xi   \frac{J_0 (\sqrt{\xi^2}) }{ 2   \sqrt{  \xi^2   +  qB  r^2 } }  =  \frac{2\pi}{4\pi} e^{-\sqrt{qB} r}   ,
 \end{equation}
   we finally obtain a explicit expression of UV divergence,
     \begin{align}
 G^{(B,L)}_{UV}  
& = \triangle G^{(B,L)}_{UV}  - \frac{2\mu e^{- \sqrt{qB} r}}{4\pi r}|_{ r \rightarrow 0} \nonumber \\
&= \triangle G^{(B,L)}_{UV} + \frac{2\mu \sqrt{qB} }{4\pi } - \frac{2\mu  }{4\pi r}|_{ r \rightarrow 0}   .   
 \end{align}

Putting all pieces together, we thus find
  \begin{align}
 & G^{(L)}_B( \mathbf{ 0},\mathbf{ 0}; \varepsilon)    \nonumber \\
 &= \triangle G^{(L)}_B (\varepsilon) + G^{(B,L)}_{RC}+\triangle G^{(B,L)}_{UV}+ \frac{2\mu \sqrt{qB} }{4\pi } - \frac{2\mu}{4\pi r}|_{ r \rightarrow 0} , \label{regGBL}
  \end{align}
where $ \triangle G^{(L)}_B (\varepsilon) $, $G^{(B,L)}_{RC} $ and $\triangle G^{(B,L)}_{UV}$  are all free of UV divergence and  given by Eq.(\ref{GBLsubtraction}), Eq.(\ref{GBLRC}) and Eq.(\ref{deltaGUV}) respectively.

\subsection{Regulated $S$-wave quantization condition}
 With explicitly isolated UV divergence in finite volume magnetic Green's function   in Eq.(\ref{regGBL}), the UV divergent terms in quantization condition given by  Eq.(\ref{unregQC})  cancel out, thus we find a regulated quantization condition
 \begin{equation}
  \cot \delta_0 (k_\varepsilon)=  \mathcal{M}^{(B,L)}_{0,0} ( \varepsilon), \label{regQC}
\end{equation}
where
\begin{align}
& \mathcal{M}^{(B,L)}_{0,0} ( \varepsilon) \nonumber \\
& = - \frac{4\pi}{2\mu k_\varepsilon}  \left [ \triangle G^{(L)}_B (\varepsilon) + G^{(B,L)}_{RC}+\triangle G^{(B,L)}_{UV} \right ] - \frac{\sqrt{qB}}{ k_\varepsilon}   . \label{magneticzetafunc}
\end{align}
The expression of $ \triangle G^{(L)}_B (\varepsilon) $, $G^{(B,L)}_{RC} $  and  $\triangle G^{(B,L)}_{UV}$   are    given by Eq.(\ref{GBLsubtraction}), Eq.(\ref{GBLRC})  and Eq.(\ref{deltaGUV})  respectively.

\subsection{L\"uscher formula at the limit of $qB \rightarrow 0$}

Using the identity given in Eq.(\ref{gammakummer}), the Eq.(\ref{GBL00}) can be rewritten as
  \begin{align}
 & G^{(L)}_B( \mathbf{ 0},\mathbf{ 0}; \varepsilon)  =  -  \frac{2\mu}{4\pi}  \frac{1}{L} \sum_{  k_{z} = \frac{2\pi n_{z}}{L}   }^{ n_{z} \in \mathbb{Z}}    \Gamma(\frac{1}{2}+ \frac{k_z^2 - 2\mu \varepsilon}{2q B})   \nonumber \\
 & \times  \sum_{n_x, n_y  }       e^{ - \frac{i q B n_x    n_y  n_q L^2 }{2}   }    e^{- \frac{qB}{4}  |  n_x n_q L \mathbf{ e}_x + n_y L \mathbf{ e}_y|^2  }      \nonumber \\
 & \times    U ( \frac{1}{2} + \frac{k_z^2 -2\mu \varepsilon}{2q B}, 1, \frac{q B}{2}  |   n_x n_q L \mathbf{ e}_x + n_y L \mathbf{ e}_y|^2 ).  
 \end{align}
 
 At the limit of $qB \rightarrow 0$, using asymptotic form of 
 \begin{align}
&  \Gamma(\frac{1}{2}+ \frac{k_z^2 - 2\mu \varepsilon}{2q B})  U ( \frac{1}{2} + \frac{k_z^2 -2\mu \varepsilon}{2q B}, 1, \frac{q B}{2}  r^2 ) \nonumber \\
& \stackrel{q B \rightarrow 0}{\rightarrow}   i \pi H_0^{(1)}( \sqrt{2\mu \varepsilon - k_z^2 } r) + \mathcal{O} (q B),
 \end{align}
 and also taking $n_q=1$, thus one find
  \begin{align}
 & G^{(L)}_B( \mathbf{ 0},\mathbf{ 0}; \varepsilon)  \stackrel{q B \rightarrow 0}{\rightarrow}   -    \frac{2\mu}{L} \sum_{  k_{z} = \frac{2\pi n_{z}}{L}   }^{ n_{z} \in \mathbb{Z}}       \nonumber \\
 & \times  \sum_{n_x, n_y  }           \frac{ i}{4}   H_0^{(1)}( \sqrt{2\mu \varepsilon - k_z^2 }   |   n_x   L \mathbf{ e}_x + n_y L \mathbf{ e}_y| ) + \mathcal{O} (q B) .  
 \end{align}
 
 Next, using identity
 \begin{align}
 & -  \sum_{n_x, n_y  }           \frac{ i}{4}   H_0^{(1)}( \sqrt{2\mu \varepsilon - k_z^2 }   |   n_x   L \mathbf{ e}_x + n_y L \mathbf{ e}_y| ) \nonumber \\
 & =  \frac{1}{L^2} \sum_{  k_{x,y} = \frac{2\pi n_{x,y}}{L}  , n_{x,y} \in \mathbb{Z}}   \frac{1}{ 2\mu \varepsilon - k_z^2  - k_x^2 - k_y^2 },
 \end{align}
 one thus can easily show that 
\begin{equation}
G^{(L)}_B(\mathbf{ 0},\mathbf{ 0}; \varepsilon ) \stackrel{q B \rightarrow 0}{\rightarrow}  G^{(L)}_0(\mathbf{ 0} ,k_\varepsilon  ) + \mathcal{O} (q B) ,
\end{equation}
where
\begin{equation}
 G^{(L)}_0(\mathbf{ r},k )  = \frac{2\mu}{L^{3}} \sum_{\mathbf{ p} = \frac{2 \pi}{L} \mathbf{ n} , \mathbf{ n} \in \mathbb{Z}^{3}  } \frac{e^{i \mathbf{ p} \cdot \mathbf{ r} }}{k^{2} - \mathbf{ p}^{2}}    . 
\end{equation}
Hence, finite volume magnetic zeta function $\mathcal{M}^{(B,L)}_{0,0} ( \varepsilon) $ at the limit of $qB \rightarrow 0$  is given by
\begin{equation}
\mathcal{M}^{(B,L)}_{0,0} ( \varepsilon)  \stackrel{q B \rightarrow 0}{\rightarrow}   \mathcal{M}^{(L)}_{ 0,0 }  ( k_\varepsilon )  + \mathcal{O} (q B) , \label{qBzeroperturbation}
\end{equation}
which is consistent with perturbation result  given in  Ref.~\cite{Detmold:2004qn}. 
The  $ \mathcal{M}^{(L)}_{ 0,0 }  ( k_\varepsilon )$  denotes the
 regular finite volume zeta function, see \cite{Luscher:1990ux,Rummukainen:1995vs,Guo:2012hv}, and is defined by
\begin{align}
  &  \mathcal{M}^{(L)}_{ 0,0 }  ( k_\varepsilon )  
    =- \frac{4\pi}{ k_\varepsilon L^{3}}  \sum_{\mathbf{ p} = \frac{2\pi}{L} \mathbf{ n} , \mathbf{ n} \in \mathbb{Z}^{3}}       \frac{e^{\frac{k_\varepsilon^{2} - \mathbf{ p}^{2}}{\Lambda^2}}  }{ k_\varepsilon^{2}-  \mathbf{ p}^{2}   }    \nonumber \\
 & +     \frac{ \Lambda}{  k_\varepsilon \sqrt{\pi}} \left [ \sum_{\mathbf{ n} \in \mathbb{Z}^{3}}^{\mathbf{ n} \neq 0}   \int_{1}^{\infty} d \xi      e^{- \frac{( \mathbf{ n}L \Lambda  \xi)^{2}  }{4} + \frac{k^{2}  }{ (\Lambda \xi)^{2}} }    +       \sum_{n=0}^{\infty} \frac{ ( \frac{ k_\varepsilon^{2}}{ \Lambda^2})^{n}}{n!(2n-1)} \right ], \label{regularzetafunc}
\end{align}
where $\Lambda$ is an arbitrary  UV regulator.

\begin{figure}
\begin{center}
\includegraphics[width=0.9\textwidth]{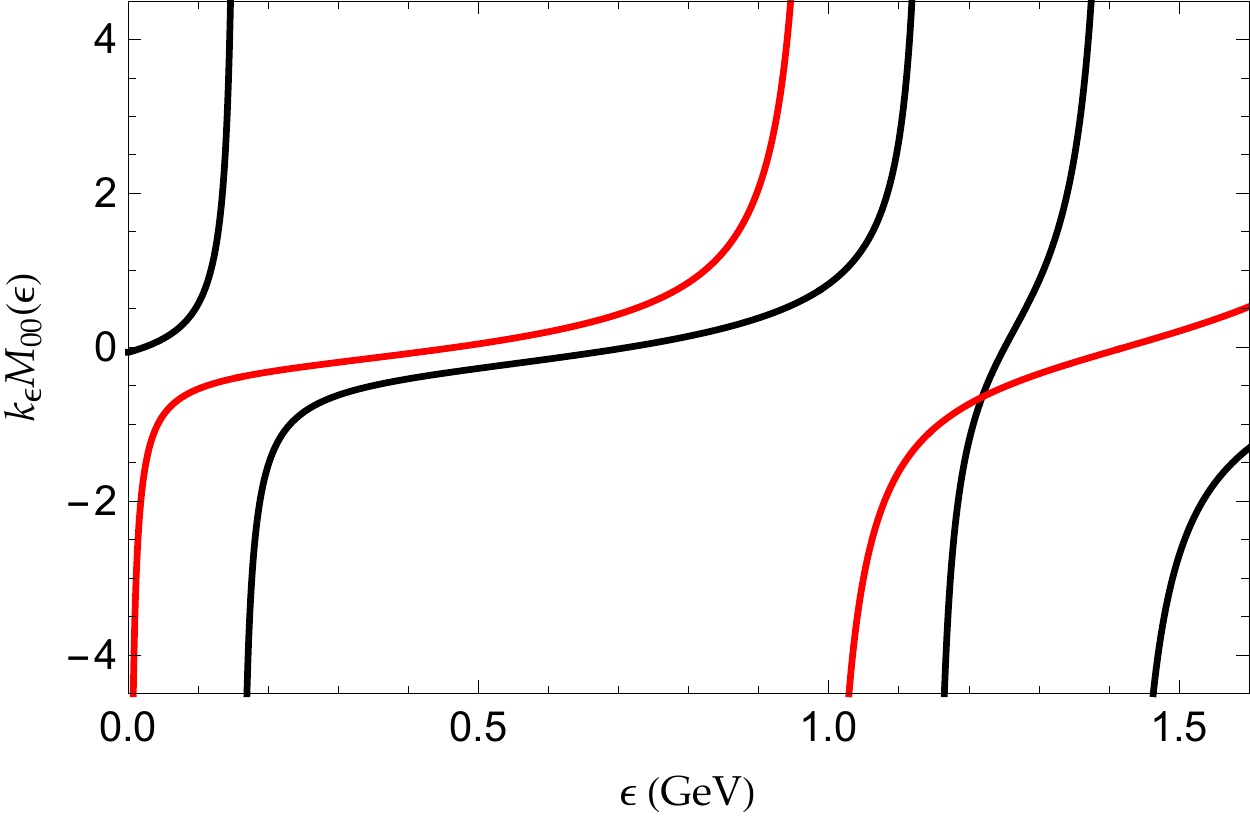}
\includegraphics[width=0.9\textwidth]{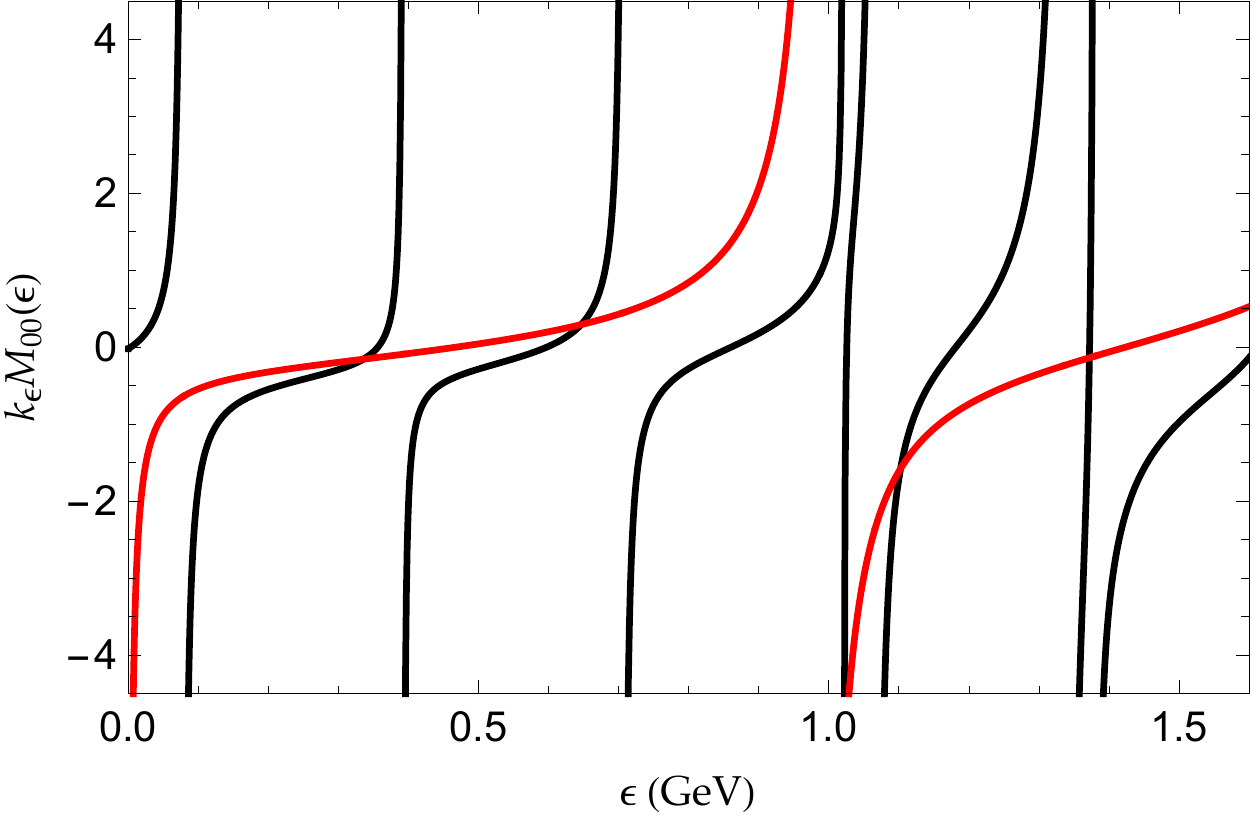}
\caption{  The comparison of the finite volume magnetic zeta function $k_\varepsilon \mathcal{M}^{(B,L)}_{0,0} (\varepsilon) $ (black solid) and regular finite volume zeta function $k \mathcal{M}^{(L)}_{0,0}  (k)$ (red solid) in Eq.(\ref{magneticzetafunc}) and Eq.(\ref{regularzetafunc}) respectively.  The parameters are chosen as: $\mu=0.2 \mbox{GeV}$,  $L=10 \mbox{GeV}^{-1}$,  $n_p=1$ and $n_q=1,2$ in upper and lower panels respectively.   }\label{ML10nqplot}
\end{center}
\end{figure}

The comparison of  the finite volume magnetic zeta function $\mathcal{M}^{(B,L)}_{0,0} (\varepsilon) $  and regular finite volume zeta function $\mathcal{M}^{(L)}_{0,0}  (k)$ are shown in Fig.~\ref{ML10nqplot}.  The splitting of energy levels are illustrated  in the comparison of the curves of $\mathcal{M}^{(B,L)}_{0,0} (\varepsilon) $ in   upper and lower panels with $n_q=1$    and $n_q =2$ respectively,   the number of $\mathcal{M}^{(B,L)}_{0,0} (\varepsilon) $ curves double as the value of $n_q$ is doubled.

\section{Summary}\label{summ}

A formalism for describing charged spinless bosons interaction in both a finite volume and   a magnetic field is presented in this work. We   show that for a short-range potential,   a L\"uscher formula-like relation   that relates discrete energy spectrum to scattering phase shifts can be obtained.  The regularization of UV divergence is worked out explicitly for $S$-wave contribution, the regulated $S$-wave quantization condition may be useful for the LQCD study of charged boson system, such as $\pi^+$ or $K^+$ system.  In finite volume and in magnetic field,   translation symmetry of system is only preserved when the magnetic flux, $\Phi_B=qBL^2$,    is given by $2\pi$ multiplied by a rational number $n_p/n_q$ where $n_p$ and $n_q$ are relatively prime integers. The presence of magnetic field thus result in the splitting of      energy level    into $n_q$ sub-energy levels.

\acknowledgments

P.G. acknowledges support from the Department of Physics and Engineering, California State University, Bakersfield, CA. This research (PG) was also supported in part by the National Science Foundation (US) under Grant No. NSF PHY-1748958.

\appendix

\section{Two charged bosons in a uniform magnetic field}\label{appenddynamics}
 
 The  dynamics of two charged non-relativistic  identical bosons in a uniform magnetic field is described by Schr\"odinger equation,
 \begin{equation}
\left  [ E+  \sum_{i =1}^2  \frac{\left (\nabla_i+ i e \mathbf{ A} (\mathbf{ x}_i) \right )^2}{2m} - V(\mathbf{ x}_1 -\mathbf{ x}_2)\right ] \Psi_E(\mathbf{ x}_1,\mathbf{ x}_2) =0,
\end{equation} 
 where  $m$ is the mass of identical bosons. $\mathbf{ x}_i$ denotes the position of i-th particle, and the short-range interaction between two particles is represented by $V(\mathbf{ x}_1 -\mathbf{ x}_2)$.  $ \mathbf{ A} (\mathbf{ x}_i)$ stands for the vector potential of uniform magnetic field. 
 
Throughout the entire   work, the uniform magnetic field is assumed along the $z$-axis, $\mathbf{ B} = B  \mathbf{ e}_z $, the vector potential in Landau gauge is used,
 \begin{equation}
    \mathbf{ A} (\mathbf{ x}_i) =  B(0, x_i, 0).
 \end{equation} 
 The solutions of Schr\"odinger equation in other gauges are obtained by a gauge transformation through a scalar field, $\chi (\mathbf{ x}_i)$,
 \begin{equation}
   \mathbf{ A} (\mathbf{ x}_i) \rightarrow   \mathbf{ A} (\mathbf{ x}_i)- \nabla \chi (\mathbf{ x}_i),  
 \end{equation}
 and
  \begin{equation}
     \Psi_E(\mathbf{ x}_1,\mathbf{ x}_2) \rightarrow e^{i  e \sum_{i=1}^2 \chi (\mathbf{ x}_i )} \Psi_E(\mathbf{ x}_1,\mathbf{ x}_2) .
 \end{equation}

\subsection{Separation of center of mass and relative motions}\label{separation} 
 The center of mass motion (CM) and relative motion of two particles can be separated by introducing CM and relative coordinates respectively
 \begin{equation}
 \mathbf{ R} = \frac{\mathbf{ x}_1 + \mathbf{ x}_2}{2}, \ \ \ \ \mathbf{ r}= \mathbf{ x}_1 -\mathbf{ x}_2.
 \end{equation}
 Therefore, the Hamiltonian has a separable form and the total two particles wave function is given by the product of CM and relative wave functions,
 \begin{equation}
 \Psi_E(\mathbf{ x}_1,\mathbf{ x}_2) = \Phi_{E-\varepsilon}(\mathbf{ R}) \psi_{  \varepsilon}(\mathbf{ r}) ,
 \end{equation}
 where CM wave function, $  \Phi_{ E- \varepsilon}(\mathbf{ R}) $, and relative wave function,  $ \psi_{  \varepsilon}(\mathbf{ r}) $, satisfy  Schr\"odinger equations respectively,
 \begin{equation}
 \left  [ (E- \varepsilon) +   \frac{\left (\nabla_{\mathbf{ R}}+ i  Q \mathbf{ A} (\mathbf{ R}) \right )^2}{2 M}  \right ]  \Phi_{ E- \varepsilon}(\mathbf{ R})  =0,
 \end{equation}
 and
  \begin{equation}
 \left  [   \varepsilon +    \frac{\left (\nabla_{\mathbf{ r}}+ i q \mathbf{ A} (\mathbf{ r}) \right )^2}{2\mu} - V(\mathbf{ r} )\right ]  \psi_{  \varepsilon}(\mathbf{ r}) =0.
 \end{equation}
 The  total and reduced mass of two particles are $$M=2 m \ \ \mbox{and} \ \ \mu = \frac{m}{2}$$ respectively, and similarly $$Q=2 e \ \  \mbox{and} \ \  q=\frac{e}{2}$$ are total and reduced charges respectively.

\subsection{Magnetic translation group and magnetic periodic boundary condition}\label{MTranslation} 
 Now, let's consider  putting charged particles in a periodic cubic box with size $L$, and interaction between two particles is also periodic,
 \begin{equation}
 V(\mathbf{ r} + \mathbf{ n} L)= V(\mathbf{ r} ), \ \ \ \ \mathbf{ n} \in \mathbb{Z}^3.
 \end{equation}
 Without magnetic field,   the discrete translation symmetry of system in finite volume yields the conserved   total momentum of system with   discrete values: $$\mathbf{ P} = \frac{2\pi \mathbf{ n}}{L},\ \ \ \  \mathbf{ n} \in \mathbb{Z}^3.$$ 
 In magnetic field, translation symmetry is explicitly broken by  position dependent vector potential $\mathbf{ A}(\mathbf{ r} )$, 
\begin{equation}
\mathbf{ A} (\mathbf{ r} +  n L \mathbf{ e}_x  )   = \mathbf{ A} (\mathbf{ r})  +  B   n L \mathbf{ e}_y  , \ \ \ \ n \in \mathbb{Z},
\end{equation}
hence, the Hamiltonian is no longer invariant under translation operation in general. Fortunately, the magnetic translation operators can be introduced, see Refs.~\cite{PhysRev.133.A1038,PhysRev.134.A1602,yoshioka2002quantum}. For instance, the magnetic translation operator for relative motion can be defined by
\begin{equation}
 \hat{T}_\mathbf{ r} (\mathbf{ n} L) = e^{ i ( - i \nabla_\mathbf{ r} + q \mathbf{ A} (\mathbf{ r} ) - q \mathbf{ B} \times \mathbf{ r}  ) \cdot \mathbf{ n} L}, \label{Trel}
\end{equation}
where $ e^{ i ( - i \nabla_\mathbf{ r}   ) \cdot \mathbf{ n} L}$ is pure translation operator, and
\begin{equation}
e^{ i ( - i \nabla_\mathbf{ r}   ) \cdot \mathbf{ n} L} \psi_\varepsilon(\mathbf{ r}) = \psi_\varepsilon(\mathbf{ r} + \mathbf{ n} L).
\end{equation}
So that 
\begin{equation}
\hat{T}_\mathbf{ r} (\mathbf{ n} L)   \psi_\varepsilon(\mathbf{ r}) = e^{  i  q \left ( \mathbf{ A} (\mathbf{ r} ) - \mathbf{ B} \times \mathbf{ r} \right )   \cdot \mathbf{ n} L} \psi_\varepsilon(\mathbf{ r} + \mathbf{ n} L), \label{TopPsi}
\end{equation}
and  $\hat{T}_\mathbf{ r}$ commutes with Hamiltonian, $$[\hat{T}_{\mathbf{ r}}, \hat{H}]=0,$$ which leaves Hamiltonian invariant.  However, the magnetic translation operators do not commute with each other in general,
\begin{equation}
\hat{T}_\mathbf{ r} ( n_x L \mathbf{ e}_x)  \hat{T}_\mathbf{ r} ( n_y L \mathbf{ e}_y) =e^{- i q B L^2 n_x  n_y }  \hat{T}_\mathbf{ r} ( n_y L \mathbf{ e}_y) \hat{T}_\mathbf{ r} ( n_x L \mathbf{ e}_x) ,
\end{equation}
 where $(n_x, n_y) \in \mathbb{Z}$.

 As shown in Refs.~\cite{PhysRev.133.A1038,PhysRev.134.A1602}, when the values of $q B$ are taken as
 \begin{equation}
 q B = \frac{1}{2} e B = \frac{2\pi}{L^2} \frac{n_p}{n_q},
 \end{equation}
 where $n_p$ and $n_q$ are integers that are relatively prime. The magnetic translation operators with enlarged unit cell formed by increased size of $n_q L$ in $\mathbf{ e}_x$ direction  thus commute with each other,
 \begin{equation}
\left [ \hat{T}_\mathbf{ r} ( n_x  (n_q L) \mathbf{ e}_x) ,  \hat{T}_\mathbf{ r} ( n_y L \mathbf{ e}_y) \right  ] =0.
 \end{equation}
 Therefore magnetic translation operators with enlarged magnetic unit box that is defined by $$n_q L \mathbf{e}_x \times L \mathbf{e}_y \times L \mathbf{e}_z$$  form a discrete group that are commonly referred as magnetic translation group.

 The  translation operator for two charged particles can be introduced by
 \begin{equation}
 \hat{T}_{\mathbf{ x}_1,\mathbf{ x}_2} (\mathbf{ n}_{1B} L , \mathbf{ n}_{2B} L)  =  \hat{T}_{\mathbf{ x}_1} (\mathbf{ n}_{1B} L)   \hat{T}_{\mathbf{ x}_2} (\mathbf{ n}_{2B} L) , 
 \end{equation}
 where
 \begin{equation}
  \hat{T}_{\mathbf{ x}_{i} } (\mathbf{ n} L) = e^{ i ( - i \nabla_{\mathbf{ x}_i} + e \mathbf{ A} (\mathbf{ x}_i ) - e \mathbf{ B} \times \mathbf{ x}_i  ) \cdot \mathbf{ n} L}. \label{Txiop}
 \end{equation}
  Both $\mathbf{ n}_{1B} L$ and $\mathbf{ n}_{2B} L$ are defined in enlarged magnetic unit box,
 \begin{equation}
 \mathbf{ n}_{i B} L = n_{i x} (n_q L) \mathbf{ e}_x + n_{i y} L \mathbf{ e}_y + n_{i z} L \mathbf{ e}_z , \ \  n_{i x,iy,iz} \in \mathbb{Z}.
 \end{equation}
 We may rewrite two particles translation operator in terms of CM and relative motion quantities,  
  \begin{equation}
 \hat{T}_{\mathbf{ x}_1,\mathbf{ x}_2} (\mathbf{ n}_{1B} L , \mathbf{ n}_{2B} L)  =  \hat{T}_{\mathbf{ r}} (\mathbf{ n}_{1B} L -\mathbf{ n}_{2B} L)   \hat{T}_{\mathbf{ R}} ( \frac{ \mathbf{ n}_{1B}+\mathbf{ n}_{2B} }{2} L) ,
 \end{equation}
 where $\hat{T}_{\mathbf{ r}} $ is defined in Eq.(\ref{Trel}), and
 \begin{equation}
  \hat{T}_\mathbf{ R} (\frac{\mathbf{ n}_B  L}{2}) = e^{ i ( - i \nabla_\mathbf{ R} + Q \mathbf{ A} (\mathbf{ R} ) - Q \mathbf{ B} \times \mathbf{ R}  ) \cdot \frac{ \mathbf{ n}_B L}{2}}.  \label{TCM}
 \end{equation}
 Note that $$QB =2 e B = \frac{2\pi}{(\frac{L}{2})^2} \frac{n_p}{n_q},$$ the translation operation of CM motion may be considered as motion of composite charge particle with total charges of $Q=2e$  in a periodic box with size of  $L/2$.

 The magnetic translation invariance of system   yields
 \begin{equation}
 \hat{T}_{\mathbf{ x}_1,\mathbf{ x}_2} (\mathbf{ n}_{1B} L , \mathbf{ n}_{2B} L)   \Psi_E(\mathbf{ x}_1,\mathbf{ x}_2) =\Psi_E(\mathbf{ x}_1,\mathbf{ x}_2) .
 \end{equation}
 Using Eq.(\ref{Txiop}),   the boundary conditions for two particles   in both finite volume and   a uniform magnetic field is given by,
 \begin{align}
 &    \Psi_E(\mathbf{ x}_1+\mathbf{ n}_{1B} L,\mathbf{ x}_2+\mathbf{ n}_{2B} L)  \nonumber \\
&    = e^{-  i  e \sum_{i=1}^2 \left ( \mathbf{ A} (\mathbf{ x}_i )  -\mathbf{  B} \times \mathbf{ x}_i \right )  \cdot \mathbf{ n}_{iB} L}      \Psi_E(\mathbf{ x}_1,\mathbf{ x}_2) .
 \end{align}
 In terms of CM and relative wave functions, we have 
  \begin{align}
&   \frac{  e^{  i  Q ( \mathbf{ A} (\mathbf{ R} )  -  \mathbf{ B} \times \mathbf{ R} )  \cdot  \frac{  \mathbf{ n}_{1B} +  \mathbf{ n}_{2B} }{2} L}   \Phi_{E-\varepsilon}(\mathbf{ R} + \frac{\mathbf{ n}_{1B}  + \mathbf{ n}_{2B} }{2} L )}{ \Phi_{E-\varepsilon}(\mathbf{ R}) }    \nonumber \\
&   =  \frac{   \psi_{  \varepsilon}(\mathbf{ r})  }{e^{  i  q ( \mathbf{ A} (\mathbf{ r} ) -  \mathbf{ B} \times \mathbf{ r} )   \cdot (\mathbf{ n}_{1B}  - \mathbf{ n}_{2B} ) L } \psi_{  \varepsilon}(\mathbf{ r} + (\mathbf{ n}_{1B}  - \mathbf{ n}_{2B} ) L)  }. \label{bc}
 \end{align}

The separable form of CM motion and relative motion  in   Eq.(\ref{bc}) suggests that both sides  must be equal to a phase factor that is independent of both CM and relative coordinates.  It allows us to introduce an arbitrary parameter  $\mathbf{ P}_B$  that is associated to pure translation operator,   the phase factor may be chosen having   form of
$$e^{i  \mathbf{ P}_B \cdot  \frac{  \mathbf{ n}_{1B} +  \mathbf{ n}_{2B} }{2} L}.$$ 
Hence the CM wave function   satisfies Bloch type magnetic periodic boundary condition,
  \begin{equation}
      \Phi_{E-\varepsilon}(\mathbf{ R} + \frac{ \mathbf{ n}_B  L }{2}  )     = e^{  -i \left (  \mathbf{ P}_B  + QB  R_y \mathbf{ e}_x  \right )   \cdot     \frac{ \mathbf{ n}_B  L}{2} }      \Phi_{E-\varepsilon}(\mathbf{ R})   .
 \end{equation} 
 The boundary condition for relative wave function is   given by
   \begin{equation}
       \psi_{  \varepsilon}(\mathbf{ r} +  \mathbf{ n}_B  L)    = e^{  i    \frac{ \mathbf{ P}_B  }{2}    \cdot  \mathbf{ n}_B L }   e^{  -i   q B r_y \mathbf{ e}_x   \cdot  \mathbf{ n}_B L }   \psi_{  \varepsilon}(\mathbf{ r}) ,
 \end{equation}
 where we have also assumed 
 \begin{equation}
 e^{i  \mathbf{ P}_B \cdot   \mathbf{ n}_{B}   L}=1,
 \end{equation}
 thus
 $$\mathbf{ P}_B =2\pi  \left (  \frac{ n_x }{n_q L} \mathbf{ e}_x+\frac{ n_y }{  L} \mathbf{ e}_y+\frac{ n_z }{  L} \mathbf{ e}_z \right ),  \ \  n_{x,y,z} \in \mathbb{Z}.$$ Although $\mathbf{ P}_B$  resemble  the  total momentum of system in   absence of magnetic field,  $\mathbf{ P}_B$  is  not a conserved quantity in magnetic field. In fact, the conserved quantity can be identified as pseudo-momentum, see e.g. Ref.~\cite{yoshioka2002quantum},
\begin{equation}
\mathbf{ K}_\mathbf{ R}= \mathbf{ P}_B  +     Q ( \mathbf{ A} (\mathbf{ R} )  -  \mathbf{ B} \times \mathbf{ R} ),
\end{equation}
which is associated to the generator of magnetic translation operator for CM motion,
 \begin{equation}
  \hat{T}_\mathbf{ R} (\frac{\mathbf{ n}_B  L}{2}) = e^{ i  \mathbf{ \hat{K}}_\mathbf{ R} \cdot  \frac{\mathbf{ n}_B L}{2}}, \ \ \ \ [ \mathbf{ \hat{K}}_\mathbf{ R} , \hat{H}]=0.
  \end{equation}

\subsection{CM motion solutions}\label{CMsolution} 
 The CM motion of two charged bosons in a uniform magnetic field is described by
  \begin{equation}
 \hat{H}_\mathbf{ R}    \Phi_{ E_R }(\mathbf{ R})  = E_R  \Phi_{ E_R }(\mathbf{ R})  ,
 \end{equation}
 where
 \begin{equation}
 \hat{H}_\mathbf{ R} =  - \frac{   1}{2 M} \left [  \partial_{R_x}^2 + (   \partial_{R_y}  +  i Q  B R_x )^2 +\partial_{R_z}^2 \right ] ,
 \end{equation}
 and $ \Phi_{ E_R  } $ must satisfies boundary condition
   \begin{equation}
      \Phi_{E_R}(\mathbf{ R} + \frac{ \mathbf{ n}_B  L }{2}  )     = e^{  -i \left ( \mathbf{ P}_B  + Q B  R_y \mathbf{ e}_x  \right )   \cdot     \frac{ \mathbf{ n}_B  L}{2} }      \Phi_{E_R}(\mathbf{ R})   .
 \end{equation} 
 The solution that satisfies magnetic periodic boundary condition can be found in \cite{PhysRev.133.A1038},
 \begin{align}
& \Phi_{ E_{R , n} }(\mathbf{ R})   \nonumber \\
& = \frac{2}{L}  \sum_{k_y = \frac{4 \pi n_y n_p}{L} - P_{By}}^{n_y \in \mathbb{Z}} \phi_n (R_x + \frac{k_y }{QB}) e^{i k_y (R_y + \frac{P_{Bx}}{QB} )} e^{-i P_{Bz} R_z}  ,
 \end{align}
 where $\phi_n$ is eigen-solution of 1D harmonic oscillator potential,
 \begin{equation}
  - \frac{   1}{2 M}  \left [  \partial_{R_x}^2  - (QB)^2 R_x^2  \right ] \phi_n(R_x) = \frac{Q B}{M} (n+ \frac{1}{2}) \phi_n(R_x) .
 \end{equation}
 The eigen-energy of CM motion is given by 
 \begin{equation}
 E_{R,n} = \frac{Q B}{M} (n+ \frac{1}{2}) + \frac{P_{Bz}^2}{2M}, \ \ n=0,1,2,\cdots, \label{ECM}
 \end{equation}
  and analytic expression of $\phi_n$ is
 \begin{equation}
  \phi_n(R_x)  = \frac{1}{\sqrt{2^n n!}} \left ( \frac{QB}{\pi} \right )^{\frac{1}{4}} e^{- \frac{QB}{2} R_x^2} H_n (\sqrt{QB} R_x).
 \end{equation}

\subsection{Relative motion and finite volume Lippmann-Schwinger equation}\label{relativeLSeq} 
 The relative motion of two charged particles in a uniform magnetic field is described by
  \begin{equation}
   \left ( \hat{H}_\mathbf{ r} + V(\mathbf{ r} ) \right )  \psi_{  \varepsilon}(\mathbf{ r}) = \varepsilon  \psi_{  \varepsilon}(\mathbf{ r}) , \label{Scheqrel}
 \end{equation}
 where
  \begin{equation}
 \hat{H}_\mathbf{ r} =  -    \frac{   1}{2 \mu} \left [  \partial_{r_x}^2 + (   \partial_{r_y}  +  i q B r_x )^2 +\partial_{r_z}^2 \right ]  , \label{Hrel}
 \end{equation}
 and $  \psi_{  \varepsilon} $ must satisfies magnetic periodic  boundary condition
   \begin{equation}
       \psi_{  \varepsilon}(\mathbf{ r} +  \mathbf{ n}_B  L)    = e^{  i    \frac{ \mathbf{ P}_B  }{2}    \cdot  \mathbf{ n}_B L }   e^{  -i   q B r_y \mathbf{ e}_x   \cdot  \mathbf{ n}_B L }   \psi_{  \varepsilon}(\mathbf{ r}) . \label{Wavbdc}
 \end{equation}

The integral representation of Schr\"odinger equation (\ref{Scheqrel}) and magnetic periodic boundary condition in Eq.(\ref{Wavbdc}) together  is given by finite volume Lippmann-Schwinger equation,
\begin{equation}
\psi_{  \varepsilon}(\mathbf{ r})  = \int_{L_B^3} d \mathbf{ r}' G^{(L)}_B(\mathbf{ r},\mathbf{ r}'; \varepsilon) V(\mathbf{ r}')\psi_{  \varepsilon}(\mathbf{ r}')  , \label{LSB}
\end{equation}
 where $L_B^3$ stands for the volume of magnetic unit box defined by unit vectors $$n_q L \mathbf{ e}_x \times L \mathbf{ e}_y \times L \mathbf{ e}_z,$$ and 
 \begin{equation}
  \int_{L_B^3} d \mathbf{ r}'  = \int_{- \frac{n_q L}{2}}^{\frac{n_q L}{2}}  d r_x'   \int_{- \frac{  L}{2}}^{\frac{  L}{2}}  d r'_y  \int_{- \frac{  L}{2}}^{\frac{  L}{2}} d r'_z.
  \end{equation}
 The finite volume magnetic Green's function $G^{(L)}_B$  also must satisfy the   magnetic periodic boundary condition,
 \begin{align}
& G^{(L)}_B(\mathbf{ r},\mathbf{ r}'; \varepsilon) \nonumber \\
& = e^{ - i    \frac{ \mathbf{ P}_B  }{2}    \cdot  \mathbf{ n}_B L }   e^{  i   q B r_y \mathbf{ e}_x   \cdot  \mathbf{ n}_B L } G^{(L)}_B(\mathbf{ r}+  \mathbf{ n}_B  L,\mathbf{ r}'; \varepsilon)   \nonumber \\
 &= e^{  i    \frac{ \mathbf{ P}_B  }{2}    \cdot  \mathbf{ n}_B L }   e^{ - i   q B r'_y \mathbf{ e}_x   \cdot  \mathbf{ n}_B L }  G^{(L)}_B(\mathbf{ r},\mathbf{ r}'+  \mathbf{ n}_B  L; \varepsilon) .
 \end{align}
 The magnetic periodic boundary conditions  and  Eq.(\ref{LSB}) suggest that  $G^{(L)}_B$ is the solution of differential equation,
 \begin{align}
& \left ( \varepsilon -  \hat{H}_\mathbf{ r}   \right ) G^{(L)}_B(\mathbf{ r},\mathbf{ r}'; \varepsilon)  \nonumber \\
&= \sum_{\mathbf{ n}_B}  e^{  -i    \frac{ \mathbf{ P}_B  }{2}    \cdot  \mathbf{ n}_B L }  e^{  i   q B r_y \mathbf{ e}_x   \cdot  \mathbf{ n}_B L }  \delta(\mathbf{ r}-\mathbf{ r}' + \mathbf{ n}_B L).
 \end{align}

Now, one of the key steps therefore   is to find an analytic solution of   finite volume magnetic Green's function $G^{(L)}_B$. The $G^{(L)}_B$ can be constructed from infinite volume magnetic Green's function $G^{(\infty)}_B$, where
 \begin{align}
 & G^{(\infty)}_B(\mathbf{ r},\mathbf{ r}'; \varepsilon) =\sum_{n=0}^\infty \int_{- \infty}^\infty \frac{d k_y}{2\pi} \frac{d k_z}{2\pi} \nonumber \\
 & \times \frac{ \phi_n(r_x + \frac{k_y}{qB}) \phi^*_n(r'_x + \frac{k_y}{qB}) e^{i k_y (r_y - r'_y)}e^{i k_z (r_z - r'_z)}}{\varepsilon - \frac{qB}{\mu}(n+ \frac{1}{2}) - \frac{k_z^2}{2\mu} },
 \end{align}
 and $G^{(\infty)}_B$ satisfies equation,
  \begin{equation}
 \left ( \varepsilon -  \hat{H}_\mathbf{ r}   \right ) G^{(\infty)}_B(\mathbf{ r},\mathbf{ r}'; \varepsilon)  =  \delta(\mathbf{ r}-\mathbf{ r}'  ).
 \end{equation}
 The LS equation  (\ref{LSB})  is equivalently given in terms of  $G^{(\infty)}_B$  by
 \begin{equation}
\psi_{  \varepsilon}(\mathbf{ r})  = \int_{ -\infty}^\infty d \mathbf{ r}' G^{(\infty)}_B(\mathbf{ r},\mathbf{ r}'; \varepsilon) V(\mathbf{ r}')\psi_{  \varepsilon}(\mathbf{ r}')  .   \label{LSinf}
\end{equation}
The integration over infinite volume in Eq.(\ref{LSinf}) can be folded up to infinite sum of integration in magnetic cell,
\begin{align}
\psi_{  \varepsilon}(\mathbf{ r})  &=\sum_{\mathbf{ n}_B}   \int_{L_B^3} d \mathbf{ r}' G^{(\infty)}_B(\mathbf{ r},\mathbf{ r}' + \mathbf{ n}_B L; \varepsilon) \nonumber \\
& \times V(\mathbf{ r}'+ \mathbf{ n}_B L)\psi_{  \varepsilon}(\mathbf{ r}'+ \mathbf{ n}_B L)  .  
\end{align}
Using magnetic periodic boundary condition given in Eq.(\ref{Wavbdc}), $G^{(L)}_B$ is thus identified as
\begin{align}
& G^{(L)}_B(\mathbf{ r},\mathbf{ r}'; \varepsilon) \nonumber \\
& = \sum_{\mathbf{ n}_B}    G^{(\infty)}_B(\mathbf{ r},\mathbf{ r}' + \mathbf{ n}_B L; \varepsilon)   e^{  i    \frac{ \mathbf{ P}_B  }{2}    \cdot  \mathbf{ n}_B L }   e^{  -i   q B r'_y \mathbf{ e}_x   \cdot  \mathbf{ n}_B L }  \nonumber \\
& = \sum_{\mathbf{ n}_B}     e^{ - i    \frac{ \mathbf{ P}_B  }{2}    \cdot  \mathbf{ n}_B L }   e^{  i   q B r_y \mathbf{ e}_x   \cdot  \mathbf{ n}_B L } G^{(\infty)}_B(\mathbf{ r} + \mathbf{ n}_B L,\mathbf{ r}' ; \varepsilon)  .
\end{align}
Hence, explicitly we find
  \begin{align}
 & G^{(L)}_B(\mathbf{ r},\mathbf{ r}'; \varepsilon)  =   \sum_{n_x \in \mathbb{Z}}      e^{ - i  (  \frac{ P_{Bx}  }{2}  -q B r_y )  n_x n_q L }   \nonumber \\
 & \times  \frac{1}{L^2} \sum_{  k_{y,z} = \frac{2\pi n_{y,z}}{L} + \frac{P_{By, Bz}}{2} }^{ n_{y,z} \in \mathbb{Z}}      e^{i k_y (r_y - r'_y  )}e^{i k_z (r_z - r'_z  )} \nonumber \\
 & \times \sum_{n=0}^\infty \frac{ \phi_n(r_x +n_x n_q L + \frac{k_y}{qB}) \phi^*_n(r'_x + \frac{k_y}{qB}) }{\varepsilon - \frac{qB}{\mu}(n+ \frac{1}{2}) - \frac{k_z^2}{2\mu} }    .
 \end{align}

 The other representation of $G^{(\infty)}_B$ are given in Refs.~\cite{doi:10.1143/JPSJ.61.4314,PhysRevB.44.10280}  by
 \begin{align}
 & G^{(\infty)}_B(\mathbf{ r},\mathbf{ r}'; \varepsilon)  =e^{ - \frac{i q B}{2} (r_x + r'_x) (r_y - r'_y)} e^{- \frac{qB}{4}  | \bm{\rho} - \bm{\rho}' |^2  }  \nonumber \\
 & \times \frac{qB}{2\pi}  \int_{- \infty}^\infty   \frac{d k_z}{2\pi}   \sum_{n=0}^\infty \frac{ L_n ( \frac{qB}{2}  | \bm{\rho} - \bm{\rho}' |^2 )e^{i k_z (r_z - r'_z)}}{\varepsilon - \frac{qB}{\mu}(n+ \frac{1}{2}) - \frac{k_z^2}{2\mu} },
 \end{align}
where $$\bm{\rho} = r_x \mathbf{ e}_x + r_y \mathbf{ e}_y, \ \ \ \ \bm{\rho}' = r'_x \mathbf{ e}_x + r'_y \mathbf{ e}_y.$$ Therefore, $G^{(L)}_B$ is also given by
 \begin{align}
 & G^{(L)}_B(\mathbf{ r},\mathbf{ r}'; \varepsilon)  =  \sum_{n_x, n_y \in \mathbb{Z}}      e^{ - i  (  \frac{ P_{Bx}  }{2}  -q B r_y )  n_x n_q L }  e^{ - i  \frac{ P_{By}  }{2}    n_y L }  \nonumber \\
 & \times  e^{ - \frac{i q B}{2} (r_x + r'_x + n_x n_q L) (r_y - r'_y+ n_y L)} \nonumber \\
 & \times  e^{- \frac{qB}{4}  | \bm{\rho} - \bm{\rho}' + n_x n_q L \mathbf{ e}_x + n_y L \mathbf{ e}_y|^2  }  \frac{qB}{2\pi}   \frac{1}{L} \sum_{  k_{z} = \frac{2\pi n_{z}}{L} + \frac{P_{ Bz}}{2} }^{ n_{z} \in \mathbb{Z}}    \nonumber \\
 & \times    \sum_{n=0}^\infty \frac{ L_n ( \frac{qB}{2}  | \bm{\rho} - \bm{\rho}'  + n_x n_q L \mathbf{ e}_x + n_y L \mathbf{ e}_y|^2 )e^{i k_z (r_z - r'_z)}}{\varepsilon - \frac{qB}{\mu}(n+ \frac{1}{2}) - \frac{k_z^2}{2\mu} }.
 \end{align}

\section{Connecting bound states in a trap to infinite volume scattering state}\label{trapscatt}

 In this section, we present a   general formalism and discussion on the topic of building connections between discrete energy spectrum of bound state in a trap and infinite volume scattering  dynamics. The type of trap is not specified in follows, the typical and commonly used traps  are periodic finite box in LQCD, harmonic potential in nuclear physics, etc.

  The relative motion of  two interacting particles in a trap is described by Schr\"odinger equation
 \begin{equation}
\hat{H}_{trap} \psi^{(trap)}_\varepsilon(\mathbf{ r}) + \int_{trap} d \mathbf{ r}' V (\mathbf{ r},\mathbf{ r}')   \psi^{(trap)}_\varepsilon(\mathbf{ r}') = \varepsilon  \psi^{(trap)}_\varepsilon(\mathbf{ r}) , \label{schtrap}
 \end{equation}
 where $\hat{H}_{trap}$ stands for the trap Hamiltonian operator, the interaction between particles is described by a non-local short-range interaction $V (\mathbf{ r},\mathbf{ r}')$ in general. The effect of a trap is usually reflected by both trap Hamiltonian and boundary condition of wave function in a trap. In the case of charged particles trapped in both a periodic box and   a uniform magnetic field, $\hat{H}_{trap}  $ and boundary condition are thus given by    $ \hat{H}_\mathbf{ r} $ in Eq.(\ref{Hrel}) and  magnetic periodic boundary condition in Eq.(\ref{Wavbdc}) respectively.  The energy spectrum hence becomes discrete.

 In infinite volume,  the dynamics of two interacting particles through the same short-range interaction $V (\mathbf{ r},\mathbf{ r}')$   is given by
  \begin{equation}
 \hat{H}_{0} \psi^{( \infty)}_{\varepsilon_\infty }(\mathbf{ r}) + \int_{- \infty}^\infty d \mathbf{ r}' V (\mathbf{ r},\mathbf{ r}')   \psi^{( \infty)}_{\varepsilon_\infty}(\mathbf{ r}') = \varepsilon_\infty  \psi^{( \infty)}_{\varepsilon_\infty}(\mathbf{ r}) , \label{schinf}
 \end{equation}
 where $$ \hat{H}_{0}= - \frac{\nabla_\mathbf{ r}^2}{2\mu}. $$
 The energy spectrum of scattering solution in infinite volume  is continuous.  With a incoming plane wave, $$e^{i \mathbf{ q} \cdot \mathbf{ r}} \ \  \mbox{where} \ \  q = \sqrt{2\mu \varepsilon_\infty },$$  the asymptotic wave function of scattering states is thus described by on-shell scattering amplitudes,  
 \begin{equation}
   \psi^{( \infty)}_{\varepsilon_\infty}(\mathbf{ r})  \stackrel{ r\rightarrow \infty}{ \rightarrow  }  \sum_{ l } (2l +1) P_l( \mathbf{ \hat{q}} \cdot \mathbf{ \hat{r}})  i^l   \left [ j_l ( q r ) + i   t_l (q) h_l^{(+)} (q r) \right ] ,
 \end{equation}
 where $t_l (q) $ denotes the elastic on-shell partial wave  scattering amplitude and can be parametrized by a phase shift function  $\delta_l (q) $, 
 \begin{equation}
 t_l (q)  = \frac{1}{\cot \delta_l (q) - i}.
 \end{equation}
 We also remark that in general case, depending on the trap, the infinite volume relative energy $\varepsilon_\infty$  is related  to finite volume relative energy $\varepsilon$ by  the shared total energy. For instance, in the case of charged particles trapped in both a periodic box and   a uniform magnetic field,
 \begin{equation}
 \varepsilon_\infty + \frac{\mathbf{ P}^2}{2M} = \varepsilon + E_{R, n} =E,
 \end{equation} 
 where CM energy $E_{R, n}$ is given by Eq.(\ref{ECM}).
 
 The dynamics of particles in a trap and in infinite volume are associated by the short-range interaction potential between two particles. As far as the range of potential is far smaller than the size of  the trap, a compact  expression between phase shift of scattering states and a function, $\mathcal{M}_{l m, l' m'} (\varepsilon)$,  that reflect geometric and dynamical properties of the trap can be found, 
 \begin{equation}
\det[ \delta_{l m, l' m'} \cot \delta_l (q) - \mathcal{M}_{l m, l' m'} (\varepsilon)]=0. \label{Luscher}
\end{equation}
 In the case of finite volume in LQCD,  this relation is   well-known L\"uscher formula  \cite{Luscher:1990ux}, the matrix function $\mathcal{M}_{l m, l' m'} (\varepsilon)$ is thus zeta function.  In finite volume, the angular momentum is no longer a good quantum number due to the breaking rotation symmetry in finite volume.   In the case of harmonic trap in nuclear physics, the relation is known as BERW formula \cite{Busch98,Stetcu:2007ms, Stetcu:2010xq, Rotureau:2010uz,Rotureau:2011vf,Luu:2010hw,Yang:2016brl,Johnson:2019sps,Zhang:2019cai, Zhang:2020rhz}, where function $\mathcal{M}_{l m, l' m'}  $ becomes diagonal in angular momentum basis. The simple form of quantization condition   in Eq.(\ref{Luscher}) is the result of the presence of two distinguishable  scales: (1) short-range interaction between two particles and (2) size of trap. Hence the short-range dynamics that is described by phase shift or scattering amplitude and long-range physics due to the presence of a trap can be factorized. 
 
 The derivation of L\"uscher formula or BERW formula can be illustrated   by considering momentum space Lippmann-Schwinger equation under the assumption of separable potential, see e.g. \cite{Guo:2020iep,Guo:2020ikh,Guo:2020spn}, an example of derivation of BERW formula in momentum space is given in Appendix~\ref{singBERW}.  Here  the result is only summarized briefly     symbolically, the reaction amplitudes in both trap and infinite volume may be introduced respectively by $$\hat{t}_{trap} = - \hat{V} \hat{\psi}  \ \ \ \ \mbox{and}  \ \ \ \ \hat{t}_\infty = - \hat{V} \hat{\psi}_\infty,$$    they satisfy integral LS equations,
 \begin{equation}
\hat{ t}_{trap} (\varepsilon)= \hat{V} \hat{G}_{trap} (\varepsilon) \hat{t}_{trap} (\varepsilon),    \label{Ttrap}
 \end{equation}
 and
  \begin{equation}
\hat{ t}_{\infty}  (q)= - \hat{V} +\hat{ V} \hat{G}_{\infty}(q) \hat{t}_{\infty} (q) ,  \label{Tinf}
 \end{equation}
 where 
 \begin{equation}  
 \hat{G}_{trap} (\varepsilon)=  \frac{1}{\varepsilon - \hat{H}_{trap}}
 \end{equation} 
 and 
 \begin{equation}
 \hat{G}_{\infty} (q) =  \frac{1}{\frac{q^2}{2\mu} - \hat{H}_{0}} 
 \end{equation}
 are  Green's function in a trap and in infinite volume  respectively. Under the assumption of separable potential that is equivalent to the zero-range interaction, 
 \begin{equation}
 \widetilde{V} ( \mathbf{ k}, \mathbf{ k}')  = \sum_{l m} (k k')^l  V_l Y_{lm} (\mathbf{ \hat{k}} )Y^*_{lm} (\mathbf{ \hat{k}}' ), \label{Vmom}
 \end{equation}
   Eq.(\ref{Ttrap}) and Eq.(\ref{Tinf}) are turned into algebra equations, and can be solved analytically \cite{Guo:2020spn}. Eliminating $\hat{V}$ from two equations,  the quantization condition is thus obtained
    \begin{equation}
\det \left [\frac{1}{ \hat{ t}_{\infty}  (q)} - \hat{G}_{\infty}(q)  +  \hat{G}_{trap} (\varepsilon) \right ]=0 ,
 \end{equation}
 which is equivalent to Eq.(\ref{Luscher}).

 Though the plane wave basis in momentum space may be a  very convenient basis in finite volume and other types of traps, for the charged particles in uniform magnetic field, the momentum is no longer the conserved quantity due to the breaking translation symmetry by magnetic field. Introducing a reaction amplitude in momentum space becomes a tricky business. Therefore,  in follows,  instead of working in momentum space, we will present the general discussion of derivation of quantization condition in coordinate space under assumption of separable short-range potential again. The Fourier transform of separable potential given in Eq.(\ref{Vmom}) is 
 \begin{align}
 &V(\mathbf{ r},\mathbf{ r}') = \int \frac{d \mathbf{ k}}{(2\pi)^3}\frac{d \mathbf{ k}'}{(2\pi)^3} e^{- i \mathbf{ k} \cdot \mathbf{r }} \widetilde{V} ( \mathbf{ k}, \mathbf{ k}')e^{ i \mathbf{ k}' \cdot \mathbf{r }'} \nonumber \\
 &= \frac{ \delta(r) \delta(r') }{ (r r')^2}  \sum_{l m}   V_l  \frac{2^{2l+1} \Gamma^2 (l+ \frac{3}{2})}{  (2\pi)^3 (r r')^l } Y_{lm} (\mathbf{ \hat{r}} )Y^*_{lm} (\mathbf{ \hat{r}}' ). \label{Vcord}
 \end{align}

\subsection{Dynamical equation in a trap} 
 In the trap, the integral representation of Eq.(\ref{schtrap}) is given by the Lippmann-Schwinger equation 
\begin{align}
 \psi^{(trap)}_\varepsilon(\mathbf{ r})  &= \int_{trap} d \mathbf{ r}'' G^{(trap)} (\mathbf{ r}, \mathbf{ r}'' ; \varepsilon) \nonumber \\
 & \times   \int_{trap} d \mathbf{ r}' V (\mathbf{ r}'',\mathbf{ r}')   \psi^{(trap)}_\varepsilon(\mathbf{ r}')   , 
\end{align}  
where 
\begin{equation}
G^{(trap)} (\mathbf{ r}, \mathbf{ r}'' ; \varepsilon) = \langle\mathbf{ r}  | \frac{1}{ \varepsilon - \hat{H}_{trap}} | \mathbf{ r}'' \rangle  
\end{equation}
stands for the Green's function in a trap. The partial wave expansions
\begin{equation}
 \psi^{(trap)}_\varepsilon(\mathbf{ r})  = \sum_{ l m }  \psi^{(trap)}_{lm }(r)  Y_{lm} (\mathbf{ \hat{r}})
\end{equation}
and
\begin{align}
& G^{(trap)} (\mathbf{ r}, \mathbf{ r}'' ; \varepsilon)  \nonumber \\
& = \sum_{lm, l''m''}  Y_{lm} (\mathbf{ \hat{r}}) G_{ lm, l''m''}^{(trap)} (r, r'' ; \varepsilon) Y^*_{l''m''} (\mathbf{ \hat{r}}'')
\end{align}
 yields
\begin{align}
 \psi^{(trap)}_{lm }(r)  &= \sum_{l' m'} \int_{trap}  {r''}^2 d r'' G_{ lm, l'm'}^{(trap)} (r, r'' ; \varepsilon) \nonumber \\
 & \times   \int_{trap} {r'}^2 d r' V_{l'} (r'', r')   \psi^{(trap)}_{l'm'}(r')   . \label{LStrapPWA}
\end{align}  
Under assumption of separable potential with the form of Eq.(\ref{Vcord}), Eq.(\ref{LStrapPWA}) is turned into an algebra equation,
 \begin{align}
& \frac{\psi^{(trap)}_{lm }(r)}{r^l}  = \sum_{l' m'}  V_{l'}   \frac{2^{2l'+1} \Gamma^2 (l'+ \frac{3}{2})}{(2\pi)^3}    \nonumber \\
 & \times      \frac{G_{ lm, l'm'}^{(trap)} (r, r'' ; \varepsilon) }{   r^l  {r'' }^{l'} }  \frac{   \psi^{(trap)}_{l'm'}(r') }{{r'}^{l'}}|_{r',r''\rightarrow 0}   .
\end{align}  
hence the discrete energy spectrum   is determined by
 \begin{equation}
  \det   \left [  \frac{\delta_{lm, l'm'}}{2^{2l'+1} V_l} -   \frac{ \Gamma^2 (l'+ \frac{3}{2})}{(2\pi)^3}   \frac{G_{ lm, l'm'}^{(trap)} (r, r' ; \varepsilon) }{   r^l  {r' }^{l'} } |_{r,r'\rightarrow 0}  \right ]   =0 . \label{QCVtrap}
\end{equation}

\subsection{Infinite volume dynamical equation}  
In infinite volume,  with a incoming plane wave of $e^{i \mathbf{ q} \cdot \mathbf{ r}} $,    
 the scattering solution of two particles interaction is described by inhomogeneous  integral   Lippmann-Schwinger equation,  
\begin{align}
 & \psi^{(\infty)}_{ \varepsilon_\infty}(\mathbf{ r},\mathbf{ q})  =e^{i \mathbf{ q} \cdot \mathbf{ r}}   \nonumber \\
 & + \int_{ -\infty}^\infty d \mathbf{ r}'' G^{(\infty)} (\mathbf{ r}- \mathbf{ r}'' ; q)   \int_{-\infty}^\infty d \mathbf{ r}' V (\mathbf{ r}'',\mathbf{ r}')   \psi^{(\infty)}_{ \varepsilon_\infty}(\mathbf{ r}',\mathbf{ q})    , 
\end{align}  
where $q = \sqrt{2\mu \varepsilon_\infty}$, and the Green's function is given by
\begin{align}
& G^{(\infty)} (\mathbf{ r} - \mathbf{ r}'' ; q)  \nonumber \\
&  = \int \frac{d \mathbf{ p}}{(2\pi)^3} \frac{ e^{i \mathbf{ p} \cdot (\mathbf{ r} - \mathbf{ r}'' )}  }{\frac{q^2}{2\mu} - \frac{\mathbf{ p}^2}{2\mu}}   = - \frac{2\mu }{4\pi} i  q h_0^{(+)} (q | \mathbf{ r} - \mathbf{ r}''  |).
\end{align}
Considering partial wave expansion,
\begin{equation}
 \psi^{(\infty)}_{ \varepsilon_\infty}(\mathbf{ r},\mathbf{ q})  = \sum_{l m}  Y^*_{lm} (\mathbf{ \hat{q}}) \psi^{(\infty)}_{  l}(r,q) Y_{lm} (\mathbf{ \hat{r}}),
\end{equation}
and
\begin{align}
& G^{(\infty)} (\mathbf{ r} - \mathbf{ r}'' ; q) = \sum_{lm }  Y_{lm} (\mathbf{ \hat{r}}) G_{ l}^{( \infty)} (r, r'' ; q) Y^*_{lm} (\mathbf{ \hat{r}}'') ,\nonumber \\
& G_{ l}^{( \infty)} (r, r'' ; q)   = -2\mu i q j_l(q r_<) h_l^{(+)} (q r_>),\label{GinfPW}
\end{align}
we thus obtain
\begin{align}
&  \psi^{(\infty)}_{l }(r,q)  =4\pi i^l j_l(q r)   \nonumber \\
 & +  \int_{0}^\infty  {r''}^2 d r'' G_{ l }^{(\infty)} (r, r'' ; q)   \int_{0}^\infty {r'}^2 d r' V_{l} (r'', r')   \psi^{(\infty)}_{l}(r',q)   .  
\end{align}  
The separable potential given in Eq.(\ref{Vcord}) yields an algebra equation
\begin{align}
&  \frac{\psi^{(\infty)}_{l }(r,q) }{r^l} =4\pi i^l  \frac{j_l(q r)}{r^l}  \nonumber \\
& +  V_l \frac{2^{2l+1} \Gamma^2 (l+ \frac{3}{2}) }{(2\pi)^3}  \frac{G_{ l }^{(\infty)} (r, r'' ; q)  }{   (r r'')^l }    \frac{ \psi^{(\infty)}_{l}(r',q)  }{{r'}^l} |_{r',r''\rightarrow 0}   .  
\end{align}  
The wave function solution is thus given by
\begin{equation}
  \frac{\psi^{(\infty)}_{l }(r,q) }{r^l} =4\pi i^l  \left [ \frac{j_l(q r)}{r^l}  + i t_l(q)   \frac{  h_l^{(+)} (q r) }{   r^l }  \right ], 
\end{equation}
where the partial wave two-body scattering amplitude $t_l(q) $ is given by
\begin{equation}
 t_l(q)   =  -  \frac{ \frac{2\mu q^{2 l+1} }{(4\pi)^2}   }{ \frac{1}{ V_l } - \frac{2^{2l+1} \Gamma^2 (l+ \frac{3}{2}) }{(2\pi)^3}  \frac{G_{ l }^{(\infty)} (r', r'' ; q)  }{   (r' r'')^l }|_{r',r'' \rightarrow 0}   }    .   \label{TLinf}
\end{equation}

\subsection{Quantization condition in a trap}
Combining Eq.(\ref{QCVtrap}) and Eq.(\ref{TLinf}), and eliminating $V_l$, one thus find
 \begin{align}
  &\det   \bigg [  \delta_{lm, l'm'}       \frac{2\mu q^{2 l+1}}{t_l(q)}  \nonumber \\
  &- \delta_{lm, l'm'}  \frac{2^{2l+3} \Gamma^2 (l+ \frac{3}{2}) }{  (2\pi)}  \frac{G_{ l }^{(\infty)} (r, r' ; q)  }{   (r r')^l }|_{r,r' \rightarrow 0}      \nonumber \\
  & +  \frac{2^{2l'+3} \Gamma^2 (l'+ \frac{3}{2})}{(2\pi)}   \frac{G_{ lm, l'm'}^{(trap)} (r, r' ; \varepsilon) }{   r^l  {r' }^{l'} } |_{r,r'\rightarrow 0}  \bigg ]   =0 .  
\end{align}  
Using asymptotic form of 
\begin{align}
&   \frac{2^{2l+3} \Gamma^2 (l+ \frac{3}{2}) }{2\mu (2\pi)}  \frac{G_{ l }^{(\infty)} (r, r' ; q)  }{   (r r')^l } |_{r,r' \rightarrow 0}   \nonumber \\
&  =   - i   q^{2 l+1} - \frac{2^{2l+1} \Gamma(l+\frac{1}{2})\Gamma(l+\frac{3}{2})}{\pi} \frac{1}{{r}^{2l+1}} |_{r\rightarrow 0}, \label{asymGinfPW}
\end{align}
and also the parameterization of $$t^{-1}_l(q) = \cot \delta_l (q) - i ,$$ thus the quantization condition in a trap is indeed  given by a L\"uscher formula-like relation,
 \begin{equation}
  \det   \left  [  \delta_{lm, l'm'}        \cot \delta_l (q)  - \mathcal{M}_{lm, l'm'}  (\varepsilon)    \right  ]   =0 ,
\end{equation}  
where
 \begin{align}
  &  \mathcal{M}_{lm, l'm'}  (\varepsilon)   = -\frac{2^{2l'+3} \Gamma^2 (l'+ \frac{3}{2})}{ 2\mu q^{2l+1}  (2\pi)}   \frac{G_{ lm, l'm'}^{(trap)} (r, r' ; \varepsilon) }{   r^l  {r' }^{l'} } |_{r,r'\rightarrow 0}  \nonumber \\
  &  - \delta_{lm, l'm'}   \frac{2^{2l+1} \Gamma(l+\frac{1}{2})\Gamma(l+\frac{3}{2})}{\pi} \frac{1}{(q r)^{2l+1}} |_{r\rightarrow 0}   . \label{Mzeta}
\end{align}  
The second term in Eq.(\ref{Mzeta}) is an ultraviolet counter term that would cancel out the ultraviolet divergent term in $G_{ lm, l'm'}^{(trap)}$, ultimate result is finite and well-defined.

\section{Momentum space LS equation and particles interaction in a harmonic trap}\label{singBERW}

In this section, we present some technical details of non-relativistic spinless  particles interaction in a harmonic trap.  
 The dynamics of  non-relativistic bosonic particles interaction in a harmonic trap is described by
\begin{equation}
\left (  \hat{H}^{(ho)}  + \hat{V}  \right )
  \Psi_E^{(ho)}( \mathbf{ x}_1, \mathbf{ x}_2 )
   = E  \Psi_E^{(ho)}( \mathbf{ x}_1, \mathbf{ x}_2 ) ,
\end{equation}
 where  
 \begin{equation}
 \hat{H}^{(ho)} = \sum_{i=1}^2  \left  ( - \frac{\nabla_i^2}{2 m} + \frac{1}{2} m  \omega^2 \mathbf{ x}^2_i \right ) ,
 \end{equation}
and $\mathbf{ x}_i$  again stand for the i-th particle's position,  the $\hat{V} $ represents the interaction between two particles. $\omega$ is the angular frequency of the oscillator.  The separation of  CM    and relative motions
$$\hat{H}^{(ho)} = \hat{H}_\mathbf{ R}^{(ho)}+\hat{H}_\mathbf{ r}^{(ho)} ,$$ where 
$$\hat{H}_\mathbf{ R}^{(ho)}= -  \frac{\nabla_{\mathbf{ R}}^2  }{2 M} + \frac{1}{2} M \omega^2 \mathbf{ R}^2 $$ and $$ \hat{H}_\mathbf{ r}^{(ho)}=- \frac{\nabla_{\mathbf{ r} }^2}{2 \mu }  + \frac{1}{2} \mu \omega^2 \mathbf{ r}^2$$
yields again
$$\Psi_E^{(ho)}( \mathbf{ x}_1, \mathbf{ x}_2 )   =   \Phi^{(ho)}_{E_{R,n}} ( \mathbf{ R})   \psi_\varepsilon^{(ho)} ( \mathbf{ r}  ).$$

The CM wave function $  \Phi^{(ho)}_{E_{R,n}} ( \mathbf{ R})  $ is the solution of $3D$ harmonic oscillator potential,
\begin{equation}
\hat{H}_\mathbf{ R}^{(ho)}   \Phi^{(ho)}_{E_{R,n}} ( \mathbf{ R})   = E_{R,n}   \Phi^{(ho)}_{E_{R,n}} ( \mathbf{ R})  , 
\end{equation}
where eigen-energy is given by
\begin{equation}
  E_{R,n} = \omega(n+ \frac{3}{2}), \ \ \ \ n=0,1,2, \cdots.
\end{equation}

The relative wave function $ \psi_\varepsilon^{(ho)} ( \mathbf{ r}  )$ satisfies Lippmann-Schwinger  equation,
\begin{equation}
\psi_\varepsilon^{(ho)} ( \mathbf{ r}  ) = \int d \mathbf{ r}' G^{(ho)} (\mathbf{ r}, \mathbf{ r}';   \varepsilon)  \int d \mathbf{ r}''  V(\mathbf{ r}',\mathbf{ r}'')  \psi^{(ho)}_\varepsilon ( \mathbf{ r}''  ), 
\end{equation}
where Green's function satisfies equation,
\begin{equation}
\left (  \varepsilon- \hat{H}_\mathbf{ r}^{(ho)} \right ) G^{(ho)} (\mathbf{ r}, \mathbf{ r}'; \varepsilon)
   =  \delta(\mathbf{ r} -\mathbf{ r}').
\end{equation}
The analytic expression of Green's function in harmonic trap is given by \cite{Blinder83}
\begin{align}
& G^{(ho)} (\mathbf{ r}, \mathbf{ r}'; \varepsilon) = \sum_{lm} Y_{lm} (\mathbf{ \hat{r}}) G^{(ho)}_l (r, r'; \varepsilon) Y^*_{lm} (\mathbf{ \hat{r}}'),  \nonumber \\
& G^{(ho)}_l (r, r'; \varepsilon) = - \frac{1}{\omega (r r')^{ \frac{3}{2}}} \frac{ \Gamma (\frac{l}{2} + \frac{3}{4} - \frac{ \varepsilon}{2 \omega}) }{\Gamma(l+\frac{3}{2})} \nonumber \\
& \times \mathcal{M}_{\frac{\varepsilon}{2\omega}, \frac{l}{2} + \frac{1}{4} } ( \mu \omega r^2_{<}) \mathcal{W}_{\frac{\varepsilon}{2 \omega},\frac{l}{2} + \frac{1}{4}}  (\mu \omega r^2_{>}) , \label{GhoPW}
 \end{align}
 where $ \mathcal{M}_{a,b}(x)$ and $\mathcal{W}_{a,b}(x)$ are Whittaker functions \cite{whittaker_watson_1996}.

\subsection{Momentum space LS equation and reaction amplitude in a harmonic oscillator trap}

The reaction amplitude in a harmonic trap can be defined by
\begin{equation}
T^{(ho)}_\varepsilon (\mathbf{ k}) = -  \int d \mathbf{ r}  e^{- i \mathbf{ k} \cdot \mathbf{ r}}  \int d \mathbf{ r}''  V(\mathbf{ r}',\mathbf{ r}'')  \psi^{(ho)}_\varepsilon ( \mathbf{ r}''  ),
\end{equation}
and  $T^{(ho)}_\varepsilon(\mathbf{ k})$ satisfies momentum space LS equation,
\begin{align}
& T^{(ho)}_\varepsilon (\mathbf{ k})   \nonumber \\
&= \int \frac{ d \mathbf{ k'} }{(2\pi)^3}   \frac{d \mathbf{ k}''}{(2\pi)^3}  \widetilde{V} ( \mathbf{ k} , \mathbf{ k}' ) \widetilde{G}^{(ho)} (\mathbf{ k}', \mathbf{ k}''; \varepsilon)   T^{(ho)}_\varepsilon (\mathbf{ k}'' ) , \label{LSsingle}
\end{align}
where $\widetilde{V}$ and $\widetilde{G}^{(ho)}$ are the Fourier transform of interaction potential $V$ and Green's function $G^{(ho)}$ respectively.  In harmonic oscillator trap, rotation symmetry is intact, hence the angular momentum is still a good quantum number, the partial wave expansion of 
$$  T^{(ho)}_\varepsilon ( \mathbf{ k} )   = \sum_{lm}   T_{l}^{(ho)} (k) Y_{lm }(\mathbf{ \hat{k}}) $$ and 
\begin{equation}
\widetilde{G}^{(ho)} (\mathbf{ k}, \mathbf{ k}'; \varepsilon) = \sum_{lm} Y_{lm} (\mathbf{ \hat{k}}) \widetilde{G}^{(ho)}_l (k, k'; \varepsilon) Y^*_{lm} (\mathbf{ \hat{k}}')
\end{equation}
 thus yields
\begin{align}
& T^{(ho)}_{l} (k)   \nonumber \\
&= \int_0^\infty \frac{  {k'}^2 d  k' }{(2\pi)^3}    \frac{  {k''}^2 d  k'' }{(2\pi)^3}   \widetilde{V}_l (  k , k') \widetilde{G}^{(ho)}_l ( k', k''; \varepsilon)  T^{(ho)}_{l}( k'' ) . \label{LSsinglePW}
\end{align}

The separable potential
$$ \widetilde{V}_l (  k, k')  = (k k')^l V_l   $$
  suggests that 
  \begin{equation}
  T^{(ho)}_{l}( k ) = k^l  t^{(ho)}_{l} ,
  \end{equation}
hence the quantization condition under assumption of separable potential is given by
\begin{equation}
\frac{1}{V_l } =   \int_0^\infty \frac{  {k'}^2 d  k' }{(2\pi)^3}    \frac{  {k''}^2 d  k'' }{(2\pi)^3}   (k' k'')^l \widetilde{G}^{(ho)}_l ( k', k''; \varepsilon)   .   \label{Vsingle}
\end{equation}

\subsection{Momentum space LS equation and scattering amplitude in  infinite volume}
In infinite volume,  the   scattering amplitude is defined by
\begin{equation}
T_{\varepsilon_\infty}^{(\infty)}( \mathbf{ k} , \mathbf{ q})  = -  \int d \mathbf{ r}  e^{- i \mathbf{ k} \cdot \mathbf{ r}}  \int d \mathbf{ r}'  V(\mathbf{ r},\mathbf{ r}')  \psi^{(\infty)}_{\varepsilon_\infty} ( \mathbf{ r}' , \mathbf{ q} ) ,
\end{equation}
and it satisfies the momentum space LS equation
\begin{equation}
 T_{\varepsilon_\infty}^{(\infty)}( \mathbf{ k} , \mathbf{ q})    =  -  \widetilde{V} ( \mathbf{ k} , \mathbf{ q} )  +\int \frac{ d \mathbf{ k}' }{(2\pi)^3}  \frac{   \widetilde{V} ( \mathbf{ k} , \mathbf{ k}' )  }{ \frac{{\mathbf{ k}'}^2}{2\mu}  -\frac{q^2}{2\mu} }    T_{\varepsilon_\infty}^{(\infty)}(\mathbf{ k}', \mathbf{ q} ) .
\end{equation}
The partial wave expansion
$$  T_{\varepsilon_\infty}^{(\infty)}( \mathbf{ k} , \mathbf{ q})  = \sum_{lm}   T_{l}^{(\infty)} (k,q) Y_{lm }(\mathbf{ \hat{k}}) Y^*_{lm }(\mathbf{ \hat{q}}) $$
 yields,
\begin{equation}
  T^{(\infty)}_{l} (k,q)  = - \widetilde{V}_l (  k, q)  +  \int_0^\infty \frac{  {k'}^2 d  k' }{(2\pi)^3} \frac{  \widetilde{V}_l (  k , k')  }{ \frac{{k'}^2}{2\mu}  -\frac{q^2}{2\mu} }   T^{(\infty)}_{l}( k' ,q) .   \label{LSsinglePWinf}
\end{equation}

The assumption of separable potential again yields an analytic solution  of scattering amplitude, 
 \begin{equation}
T^{(\infty)}_{l} (q',q)  = - \frac{ (q' q)^L}{  \frac{1}{  V_l }-  \int_0^\infty \frac{  k^2 d  k }{(2\pi)^3}  \frac{ k^{2 l}   }{ \frac{k^2}{2\mu}  -\frac{q^2}{2\mu} }     }   .
\end{equation}
The on-shell partial wave scattering amplitudes $T^{(\infty)}_{l} (q,q) $, where $q=\sqrt{2\mu \varepsilon_\infty}$,  are usually parameterized by   phase shift,
\begin{equation}
T^{(\infty)}_{l} (q,q) =   \frac{(4\pi)^2}{2\mu q} \frac{1}{ \cot \delta_l (q) - i}.
\end{equation}
Therefore,  a simple relation between $ V_L$ and phase shift is obtained,
 \begin{equation}
  \frac{1}{  V_l }-  \int_0^\infty \frac{  k^2 d  k }{(2\pi)^3}  \frac{ k^{2 l}  }{ \frac{k^2}{2\mu}  -\frac{q^2}{2\mu} }      = -   \frac{2 \mu q^{2l +1}}{(4\pi)^2} [  \cot \delta_l(q) - i ].     \label{Vsingleinf}
\end{equation}

\subsection{Quantization condition in a harmonic oscillator trap}

Combing Eq.(\ref{Vsingleinf}) and Eq.(\ref{Vsingle}), we find
\begin{align}
  & \int_0^\infty \frac{  k^2 d  k }{(2\pi)^3} \frac{  {k'}^2 d  k' }{(2\pi)^3}    (kk')^l \widetilde{G}^{(ho)}_l ( k, k'; \varepsilon) \nonumber \\
  & -  \int_0^\infty \frac{  k^2 d  k }{(2\pi)^3}  \frac{k^{2 l}   }{ \frac{k^2}{2\mu}  -\frac{q^2}{2\mu} }       = -   \frac{2\mu q^{2l+1}}{(4\pi)^2} [  \cot \delta_l (q ) - i ]  .   
    \label{qcpartialwave}
\end{align}

Using asymptotic form of spherical Bessel function,
\begin{equation}
j_l (k r) \stackrel{ r\rightarrow 0}{ \rightarrow} \frac{\sqrt{\pi} (k r)^l }{2^{l+1} \Gamma(l+\frac{3}{2})},
\end{equation}
one can easily prove that
\begin{align}
 &  \int_0^\infty \frac{  k^2 d  k }{(2\pi)^3} \frac{  {k'}^2 d  k' }{(2\pi)^3}    (kk')^l \widetilde{G}^{(ho)}_l ( k, k'; \varepsilon)  \nonumber \\
 & =  \frac{1}{(4\pi)^2}      \frac{2^{2l+2} \Gamma^2(l+\frac{3}{2})}{\pi} \frac{G^{(ho)}_l (r,r'; \varepsilon)}{(r r')^l} |_{r,r' \rightarrow 0},   
 \end{align}
 and
 \begin{align}
 & \int_0^\infty \frac{  k^2 d  k }{(2\pi)^3}  \frac{k^{2 l}   }{ \frac{k^2}{2\mu}  -\frac{q^2}{2\mu} }      \nonumber \\
 &    =   \frac{1}{(4\pi)^2}      \frac{2^{2l+2} \Gamma^2(l+\frac{3}{2})}{\pi} \frac{G^{(\infty)}_l (r,r'; q )}{(r r')^l} |_{r,r' \rightarrow 0} ,
\end{align}
where the analytic expression of $G^{(ho)}_l (r,r'; \varepsilon)$  and $G^{(\infty)}_l (r,r'; q )$ are given in  Eq.(\ref{GhoPW}) and Eq.(\ref{GinfPW}) respectively.
Also using the   asymptotic form of  harmonic oscillator trap Green's function, 
\begin{align}
&      \frac{2^{2l+2}\Gamma^2(l + \frac{3}{2}) }{\pi }     \frac{G^{(ho)}_{ l} (r,r'; \varepsilon )  }{(r r')^l} |_{r,r' \rightarrow 0}  \nonumber \\
& = -   \frac{ (\mu \omega)^{l+ \frac{3}{2}}}{  \omega  }       2^{2l+2}  (-1)^{l+1}   \frac{ \Gamma (  \frac{3}{4}+\frac{l}{2} - \frac{ \varepsilon}{2 \omega}) }{\Gamma ( \frac{1}{4}-\frac{l}{2} - \frac{\epsilon_n}{2\omega} )}  \nonumber \\
&-         \frac{2^{2l+1}\Gamma(l + \frac{1}{2})  \Gamma(l + \frac{3}{2}) }{\pi }   \frac{ 2\mu   }{ r^{2l + 1}}    |_{r\rightarrow 0},  \label{asymp1}
 \end{align}
 and  asymptotic form of $G^{(\infty)}_l (r,r'; q )$ given in Eq.(\ref{asymGinfPW}), 
      the UV divergence cancel out explicitly in Eq.(\ref{qcpartialwave}),    and  the  quantization condition    is thus reduced to BERW formula,
 \begin{equation}
     \cot \delta_l (q ) -  (-1)^{l+1} (\frac{4 \mu \omega}{q^2})^{l+ \frac{1}{2}}         \frac{ \Gamma ( \frac{3}{4}+\frac{l}{2}  - \frac{ \varepsilon}{2 \omega}) }{\Gamma ( \frac{1}{4}-\frac{l}{2} - \frac{\varepsilon}{2\omega} )}   = 0     ,
\end{equation}
where $q$ and $\varepsilon$ are associated by
\begin{equation}
 \frac{q^2}{2\mu}+ \frac{ \mathbf{P}^2}{2M} = \varepsilon +  \omega (n+ \frac{3}{2} ) .
 \end{equation}

\bibliography{ALL-REF.bib}

\end{document}